\documentclass[longauth]{aa} 
\usepackage{graphicx}
\usepackage{natbib,hyperref}
\bibpunct{(}{)}{;}{a}{}{,}
\usepackage{txfonts}
\begin{document}

\title{Hint of an exocomet transit in the CHEOPS light curve of HD\,172555\thanks{This article uses data from \textit{CHEOPS} programme CH\_PR100010.}}

\titlerunning{Hint of an exocomet transit in the CHEOPS light curve of HD\,172555}
\authorrunning{Kiefer et al.}

% \author{F. Kiefer\inst{1,2}, et al.}
% \institute{\label{inst:1} Sorbonne Université, CNRS, UMR 7095, Institut d’Astrophysique de Paris, 98 bis bd Arago, 75014 Paris, France \and
%  \label{inst:2} LESIA, Observatoire de Paris, Université PSL, CNRS, Sorbonne Université, Université Paris Cité, 5 place Jules Janssen, 92195 Meudon, France}
 
\author{
F. Kiefer\inst{1,2} $^{\href{https://orcid.org/0000-0001-9129-4929}{\includegraphics[scale=0.5]{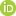}}}$\thanks{E-mail: flavien.kiefer@obspm.fr},
V. Van Grootel\inst{3} $^{\href{https://orcid.org/0000-0003-2144-4316}{\includegraphics[scale=0.5]{figures/orcid.jpg}}}$,
A. Lecavelier des Etangs\inst{4} $^{\href{https://orcid.org/0000-0002-5637-5253}{\includegraphics[scale=0.5]{figures/orcid.jpg}}}$,
Gy. M. Szabó\inst{5,6}, 
A. Brandeker\inst{7} $^{\href{https://orcid.org/0000-0002-7201-7536}{\includegraphics[scale=0.5]{figures/orcid.jpg}}}$,
C. Broeg\inst{8,9} $^{\href{https://orcid.org/0000-0001-5132-2614}{\includegraphics[scale=0.5]{figures/orcid.jpg}}}$,
A. Collier Cameron\inst{10} $^{\href{https://orcid.org/0000-0002-8863-7828}{\includegraphics[scale=0.5]{figures/orcid.jpg}}}$,
A. Deline\inst{11}, 
G. Olofsson\inst{7} $^{\href{https://orcid.org/0000-0003-3747-7120}{\includegraphics[scale=0.5]{figures/orcid.jpg}}}$,
T. G. Wilson\inst{10} $^{\href{https://orcid.org/0000-0001-8749-1962}{\includegraphics[scale=0.5]{figures/orcid.jpg}}}$,
S. G. Sousa\inst{12} $^{\href{https://orcid.org/0000-0001-9047-2965}{\includegraphics[scale=0.5]{figures/orcid.jpg}}}$,
D. Gandolfi\inst{13} $^{\href{https://orcid.org/0000-0001-8627-9628}{\includegraphics[scale=0.5]{figures/orcid.jpg}}}$,
G. Hébrard\inst{1,14}, 
Y. Alibert\inst{8} $^{\href{https://orcid.org/0000-0002-4644-8818}{\includegraphics[scale=0.5]{figures/orcid.jpg}}}$,
R. Alonso\inst{15,16} $^{\href{https://orcid.org/0000-0001-8462-8126}{\includegraphics[scale=0.5]{figures/orcid.jpg}}}$,
G. Anglada\inst{17,18} $^{\href{https://orcid.org/0000-0002-3645-5977}{\includegraphics[scale=0.5]{figures/orcid.jpg}}}$,
T. Bárczy\inst{19} $^{\href{https://orcid.org/0000-0002-7822-4413}{\includegraphics[scale=0.5]{figures/orcid.jpg}}}$,
D. Barrado\inst{20} $^{\href{https://orcid.org/0000-0002-5971-9242}{\includegraphics[scale=0.5]{figures/orcid.jpg}}}$,
S. C. C. Barros\inst{12,21} $^{\href{https://orcid.org/0000-0003-2434-3625}{\includegraphics[scale=0.5]{figures/orcid.jpg}}}$,
W. Baumjohann\inst{22} $^{\href{https://orcid.org/0000-0001-6271-0110}{\includegraphics[scale=0.5]{figures/orcid.jpg}}}$,
M. Beck\inst{11} $^{\href{https://orcid.org/0000-0003-3926-0275}{\includegraphics[scale=0.5]{figures/orcid.jpg}}}$,
T. Beck\inst{8}, 
W. Benz\inst{8,23} $^{\href{https://orcid.org/0000-0001-7896-6479}{\includegraphics[scale=0.5]{figures/orcid.jpg}}}$,
N. Billot\inst{11} $^{\href{https://orcid.org/0000-0003-3429-3836}{\includegraphics[scale=0.5]{figures/orcid.jpg}}}$,
X. Bonfils\inst{24} $^{\href{https://orcid.org/0000-0001-9003-8894}{\includegraphics[scale=0.5]{figures/orcid.jpg}}}$,
J. Cabrera\inst{25}, 
S. Charnoz\inst{26} $^{\href{https://orcid.org/0000-0002-7442-491X}{\includegraphics[scale=0.5]{figures/orcid.jpg}}}$,
Sz. Csizmadia\inst{25} $^{\href{https://orcid.org/0000-0001-6803-9698}{\includegraphics[scale=0.5]{figures/orcid.jpg}}}$,
M. B. Davies\inst{27} $^{\href{https://orcid.org/0000-0001-6080-1190}{\includegraphics[scale=0.5]{figures/orcid.jpg}}}$,
M. Deleuil\inst{28} $^{\href{https://orcid.org/0000-0001-6036-0225}{\includegraphics[scale=0.5]{figures/orcid.jpg}}}$,
L. Delrez\inst{29,3} $^{\href{https://orcid.org/0000-0001-6108-4808}{\includegraphics[scale=0.5]{figures/orcid.jpg}}}$,
O. D. S. Demangeon\inst{12,21} $^{\href{https://orcid.org/0000-0001-7918-0355}{\includegraphics[scale=0.5]{figures/orcid.jpg}}}$,
B.-O. Demory\inst{23} $^{\href{https://orcid.org/0000-0002-9355-5165}{\includegraphics[scale=0.5]{figures/orcid.jpg}}}$,
D. Ehrenreich\inst{30,31} $^{\href{https://orcid.org/0000-0001-9704-5405}{\includegraphics[scale=0.5]{figures/orcid.jpg}}}$,
A. Erikson\inst{25}, 
A. Fortier\inst{8,23} $^{\href{https://orcid.org/0000-0001-8450-3374}{\includegraphics[scale=0.5]{figures/orcid.jpg}}}$,
L. Fossati\inst{22} $^{\href{https://orcid.org/0000-0003-4426-9530}{\includegraphics[scale=0.5]{figures/orcid.jpg}}}$,
M. Fridlund\inst{32,33} $^{\href{https://orcid.org/0000-0002-0855-8426}{\includegraphics[scale=0.5]{figures/orcid.jpg}}}$,
M. Gillon\inst{29} $^{\href{https://orcid.org/0000-0003-1462-7739}{\includegraphics[scale=0.5]{figures/orcid.jpg}}}$,
M. Güdel\inst{34}, 
K. Heng\inst{23,35} $^{\href{https://orcid.org/0000-0003-1907-5910}{\includegraphics[scale=0.5]{figures/orcid.jpg}}}$,
S. Hoyer\inst{28} $^{\href{https://orcid.org/0000-0003-3477-2466}{\includegraphics[scale=0.5]{figures/orcid.jpg}}}$,
K. G. Isaak\inst{36} $^{\href{https://orcid.org/0000-0001-8585-1717}{\includegraphics[scale=0.5]{figures/orcid.jpg}}}$,
L. L. Kiss\inst{37,38}, 
J. Laskar\inst{39} $^{\href{https://orcid.org/0000-0003-2634-789X}{\includegraphics[scale=0.5]{figures/orcid.jpg}}}$,
M. Lendl\inst{11} $^{\href{https://orcid.org/0000-0001-9699-1459}{\includegraphics[scale=0.5]{figures/orcid.jpg}}}$,
C. Lovis\inst{11} $^{\href{https://orcid.org/0000-0001-7120-5837}{\includegraphics[scale=0.5]{figures/orcid.jpg}}}$,
D. Magrin\inst{40} $^{\href{https://orcid.org/0000-0003-0312-313X}{\includegraphics[scale=0.5]{figures/orcid.jpg}}}$,
P. F. L. Maxted\inst{41} $^{\href{https://orcid.org/0000-0003-3794-1317}{\includegraphics[scale=0.5]{figures/orcid.jpg}}}$,
M. Munari\inst{42} $^{\href{https://orcid.org/0000-0003-0990-050X}{\includegraphics[scale=0.5]{figures/orcid.jpg}}}$,
V. Nascimbeni\inst{40} $^{\href{https://orcid.org/0000-0001-9770-1214}{\includegraphics[scale=0.5]{figures/orcid.jpg}}}$,
R. Ottensamer\inst{43}, 
I. Pagano\inst{42} $^{\href{https://orcid.org/0000-0001-9573-4928}{\includegraphics[scale=0.5]{figures/orcid.jpg}}}$,
E. Pallé\inst{15} $^{\href{https://orcid.org/0000-0003-0987-1593}{\includegraphics[scale=0.5]{figures/orcid.jpg}}}$,
G. Peter\inst{44} $^{\href{https://orcid.org/0000-0001-6101-2513}{\includegraphics[scale=0.5]{figures/orcid.jpg}}}$,
D. Piazza\inst{45}, 
G. Piotto\inst{40,46} $^{\href{https://orcid.org/0000-0002-9937-6387}{\includegraphics[scale=0.5]{figures/orcid.jpg}}}$,
D. Pollacco\inst{35}, 
D. Queloz\inst{47,48} $^{\href{https://orcid.org/0000-0002-3012-0316}{\includegraphics[scale=0.5]{figures/orcid.jpg}}}$,
R. Ragazzoni\inst{40,46} $^{\href{https://orcid.org/0000-0002-7697-5555}{\includegraphics[scale=0.5]{figures/orcid.jpg}}}$,
N. Rando\inst{49}, 
F. Ratti\inst{49}, 
H. Rauer\inst{25,50,51} $^{\href{https://orcid.org/0000-0002-6510-1828}{\includegraphics[scale=0.5]{figures/orcid.jpg}}}$,
C. Reimers\inst{43}, 
I. Ribas\inst{17,18} $^{\href{https://orcid.org/0000-0002-6689-0312}{\includegraphics[scale=0.5]{figures/orcid.jpg}}}$,
N. C. Santos\inst{12,21} $^{\href{https://orcid.org/0000-0003-4422-2919}{\includegraphics[scale=0.5]{figures/orcid.jpg}}}$,
G. Scandariato\inst{42} $^{\href{https://orcid.org/0000-0003-2029-0626}{\includegraphics[scale=0.5]{figures/orcid.jpg}}}$,
D. Ségransan\inst{11} $^{\href{https://orcid.org/0000-0003-2355-8034}{\includegraphics[scale=0.5]{figures/orcid.jpg}}}$,
A. E. Simon\inst{8} $^{\href{https://orcid.org/0000-0001-9773-2600}{\includegraphics[scale=0.5]{figures/orcid.jpg}}}$,
A. M. S. Smith\inst{25} $^{\href{https://orcid.org/0000-0002-2386-4341}{\includegraphics[scale=0.5]{figures/orcid.jpg}}}$,
M. Steller\inst{22} $^{\href{https://orcid.org/0000-0003-2459-6155}{\includegraphics[scale=0.5]{figures/orcid.jpg}}}$,
N. Thomas\inst{8}, 
S. Udry\inst{11} $^{\href{https://orcid.org/0000-0001-7576-6236}{\includegraphics[scale=0.5]{figures/orcid.jpg}}}$,
I. Walter\inst{52} $^{\href{https://orcid.org/0000-0002-5839-1521}{\includegraphics[scale=0.5]{figures/orcid.jpg}}}$,
N. A. Walton\inst{53} $^{\href{https://orcid.org/0000-0003-3983-8778}{\includegraphics[scale=0.5]{figures/orcid.jpg}}}$
}

\institute{
\label{inst:1} Sorbonne Université, CNRS, UMR 7095, Institut d'Astrophysique de Paris, 98 bis bd Arago, 75014 Paris, France \and
\label{inst:2} LESIA, Observatoire de Paris, Université PSL, CNRS, Sorbonne Université, Université Paris Cité, 5 place Jules Janssen, 92195 Meudon, France \and
\label{inst:3} Space sciences, Technologies and Astrophysics Research (STAR) Institute, Université de Liège, Allée du 6 Août 19C, 4000 Liège, Belgium \and
\label{inst:4} Institut d'astrophysique de Paris, UMR7095 CNRS, Université Pierre \& Marie Curie, 98bis blvd. Arago, 75014 Paris, France \and
\label{inst:5} ELTE E\"otv\"os Lorandand University, Gothard Astrophysical Observatory, 9700 Szombathely, Szent Imre h. u. 112, Hungary \and
\label{inst:6} MTA-ELTE Exoplanet Research Group, 9700 Szombathely, Szent Imre h. u. 112, Hungary \and
\label{inst:7} Department of Astronomy, Stockholm University, AlbaNova University Center, 10691 Stockholm, Sweden \and
\label{inst:8} Physikalisches Institut, University of Bern, Sidlerstrasse 5, 3012 Bern, Switzerland \and
\label{inst:9} Center for Space and Habitability, Gesellsschaftstrasse 6, 3012 Bern, Switzerland \and
\label{inst:10} Centre for Exoplanet Science, SUPA School of Physics and Astronomy, University of St Andrews, North Haugh, St Andrews KY16 9SS, UK \and
\label{inst:11} Observatoire Astronomique de l'Université de Genève, Chemin Pegasi 51, CH-1290 Versoix, Switzerland \and
\label{inst:12} Instituto de Astrofisica e Ciencias do Espaco, Universidade do Porto, CAUP, Rua das Estrelas, 4150-762 Porto, Portugal \and
\label{inst:13} Dipartimento di Fisica, Universita degli Studi di Torino, via Pietro Giuria 1, I-10125, Torino, Italy \and
\label{inst:14} Observatoire de Haute-Provence, CNRS, Université d'Aix-Marseille, 04870 Saint-Michel-l'Observatoire, France \and
\label{inst:15} Instituto de Astrofisica de Canarias, 38200 La Laguna, Tenerife, Spain \and
\label{inst:16} Departamento de Astrofisica, Universidad de La Laguna, 38206 La Laguna, Tenerife, Spain \and
\label{inst:17} Institut de Ciencies de l'Espai (ICE, CSIC), Campus UAB, Can Magrans s/n, 08193 Bellaterra, Spain \and
\label{inst:18} Institut d'Estudis Espacials de Catalunya (IEEC), 08034 Barcelona, Spain \and
\label{inst:19} Admatis, 5. Kandó Kálmán Street, 3534 Miskolc, Hungary \and
\label{inst:20} Depto. de Astrofisica, Centro de Astrobiologia (CSIC-INTA), ESAC campus, 28692 Villanueva de la Ca\~nada (Madrid), Spain \and
\label{inst:21} Departamento de Fisica e Astronomia, Faculdade de Ciencias, Universidade do Porto, Rua do Campo Alegre, 4169-007 Porto, Portugal \and
\label{inst:22} Space Research Institute, Austrian Academy of Sciences, Schmiedlstrasse 6, A-8042 Graz, Austria \and
\label{inst:23} Center for Space and Habitability, University of Bern, Gesellschaftsstrasse 6, 3012 Bern, Switzerland \and
\label{inst:24} Université Grenoble Alpes, CNRS, IPAG, 38000 Grenoble, France \and
\label{inst:25} Institute of Planetary Research, German Aerospace Center (DLR), Rutherfordstrasse 2, 12489 Berlin, Germany \and
\label{inst:26} Université de Paris, Institut de physique du globe de Paris, CNRS, F-75005 Paris, France \and
\label{inst:27} Centre for Mathematical Sciences, Lund University, Box 118, 221 00 Lund, Sweden \and
\label{inst:28} Aix Marseille Univ, CNRS, CNES, LAM, 38 rue Frédéric Joliot-Curie, 13388 Marseille, France \and
\label{inst:29} Astrobiology Research Unit, Université de Liège, Allée du 6 Août 19C, B-4000 Liège, Belgium \and
\label{inst:30} Observatoire Astronomique de l'Université de Gen\`eve, Chemin Pegasi 51, CH-1290 Versoix, Switzerland \and
\label{inst:31} Centre Vie dans l'Univers, Faculté des sciences, Université de Gen\`eve, Quai Ernest-Ansermet 30, CH-1211 Gen\`eve 4, Switzerland \and
\label{inst:32} Leiden Observatory, University of Leiden, PO Box 9513, 2300 RA Leiden, The Netherlands \and
\label{inst:33} Department of Space, Earth and Environment, Chalmers University of Technology, Onsala Space Observatory, 439 92 Onsala, Sweden \and
\label{inst:34} University of Vienna, Department of Astrophysics, Türkenschanzstrasse 17, 1180 Vienna, Austria \and
\label{inst:35} Department of Physics, University of Warwick, Gibbet Hill Road, Coventry CV4 7AL, United Kingdom \and
\label{inst:36} Science and Operations Department - Science Division (SCI-SC), Directorate of Science, European Space Agency (ESA), European Space Research and Technology Centre (ESTEC), Keplerlaan 1, 2201-AZ Noordwijk, The Netherlands  \and
\label{inst:37} Konkoly Observatory, Research Centre for Astronomy and Earth Sciences, 1121 Budapest, Konkoly Thege Miklós út 15-17, Hungary \and
\label{inst:38} ELTE E\"otv\"os Lor\'and University, Institute of Physics, P\'azm\'any P\'eter s\'et\'any 1/A, 1117 \and
\label{inst:39} IMCCE, UMR8028 CNRS, Observatoire de Paris, PSL Univ., Sorbonne Univ., 77 av. Denfert-Rochereau, 75014 Paris, France \and
\label{inst:40} INAF, Osservatorio Astronomico di Padova, Vicolo dell'Osservatorio 5, 35122 Padova, Italy \and
\label{inst:41} Astrophysics Group, Keele University, Staffordshire, ST5 5BG, United Kingdom \and
\label{inst:42} INAF, Osservatorio Astrofisico di Catania, Via S. Sofia 78, 95123 Catania, Italy \and
\label{inst:43} Department of Astrophysics, University of Vienna, Tuerkenschanzstrasse 17, 1180 Vienna, Austria \and
\label{inst:44} Institute of Optical Sensor Systems, German Aerospace Center (DLR), Rutherfordstrasse 2, 12489 Berlin, Germany \and
\label{inst:45} Physics Institute, University of Bern \and
\label{inst:46} Dipartimento di Fisica e Astronomia "Galileo Galilei", Universita degli Studi di Padova, Vicolo dell'Osservatorio 3, 35122 Padova, Italy \and
\label{inst:47} ETH Zurich, Department of Physics, Wolfgang-Pauli-Strasse 2, CH-8093 Zurich, Switzerland \and
\label{inst:48} Cavendish Laboratory, JJ Thomson Avenue, Cambridge CB3 0HE, UK \and
\label{inst:49} ESTEC, European Space Agency, 2201AZ, Noordwijk, NL \and
\label{inst:50} Zentrum für Astronomie und Astrophysik, Technische Universität Berlin, Hardenbergstr. 36, D-10623 Berlin, Germany \and
\label{inst:51} Institut für Geologische Wissenschaften, Freie Universität Berlin, 12249 Berlin, Germany \and
\label{inst:52} German Aerospace Center (DLR), Institute of Optical Sensor Systems, Rutherfordstrasse 2, 12489 Berlin \and
\label{inst:53} Institute of Astronomy, University of Cambridge, Madingley Road, Cambridge, CB3 0HA, United Kingdom
}

   \date{Received on 30/09/2022; accepted on 10/12/2022}

% \abstract{}{}{}{}{}
% 5 {} token are mandatory
 
  \abstract{
   HD\,172555 is a young ($\sim$20\,Myr) A7V star surrounded by a 10\,au wide debris disk suspected to be replenished partly by collisions between large planetesimals. Small evaporating transiting bodies, that is exocomets, have also been detected in this system by spectroscopy. After $\beta$\,Pictoris, this is another example of a system possibly witnessing a phase of the heavy bombardment of planetesimals.
   In such a system, small bodies trace dynamical evolution processes. We aim to constrain their dust content by using transit photometry.
  We performed a 2-day-long photometric monitoring of HD\,172555 with the CHEOPS space telescope in order to detect shallow transits of exocomets with a typical expected duration of a few hours.
  The large oscillations in the light curve indicate that HD\,172555 is a $\delta$\,Scuti pulsating star. After removing those dominating oscillations, we found a hint of a transient absorption. If fitted with an exocomet transit model, it would correspond to an evaporating body passing near the star at 
  a distance of 6.8$\pm$1.4\,R$_\star$ (or 0.05$\pm$0.01\,au) with a radius of 2.5 km. These properties are comparable to those of the exocomets already found in this system using spectroscopy, as well as those found in the $\beta$\,Pic system. The nuclei of the Solar System's Jupiter family comets, with radii of 2--6\,km, are also comparable in size. This is the first piece of evidence of an exocomet photometric transit detection in the young system of HD\,172555.
  }
 
   \maketitle

\section{Introduction}

In the direct aftermath of planet formation, planetary systems can carry a variety of bodies with many sizes and masses: gas and dust particles within disks, small bodies that are more or less icy (asteroids and comets), and planets. How those appear and evolve together --- or one after the other --- is still an open question. Dynamical instabilities leading to planet orbit reorganisation as well as small body scattering, during or after protoplanetary disk dissipation, are thought to take place early in the life of planetary systems, and they are believed to have occurred in the young Solar System \citep{morbi2001, chambers2002, Tsiganis2005, Gomes2005, desousa2020, beibei2022}. In the Solar System, cratering impacts on the Moon are the result of the bombardment of small bodies that occurred several billion years ago \citep{ryder1990, Bottke2012, morbi2018, Marks2019, Cartwright2022}. 
How this phase relates to the configuration and properties of the individual planets in the Solar System is still under investigation, primarily because the chronology is uncertain due to the possibility of observing the Solar System only now, several billion years after these events. 
Observing young systems, such as HD\,172555 and $\beta$\,Pictoris, that may currently be undergoing a period of heavy bombardment provides a unique window into the dynamical processes at play during the first dozen million years of a planetary system.

The naked-eye star (V=4.9) HD\,172555 is a 20\,Myr-old A7V dwarf belonging to the $\beta$\,Pic moving group, whose age is well established to be within 15--25 Myr \citep{Barrado1999, Zuckerman2001, Binks2014, Mamajek2014, Miret2018}. Its principal properties are summarised in Table~\ref{tab:star}. It is the member of a wide binary system with a distant M-dwarf companion, CD-64\,1208, at $\sim$2000\,au,  or equivalently 70\arcsec\ \citep{Feigelson2006,Alonso2015}. HD\,172555 harbors a cold dust disk with an extent of up to $\sim$1000 au \citep{Nilsson2009}, and a warm 10\,au wide dusty debris disk close to edge-on at the inclination  $I_\text{disk}$$\sim$$75^\circ$  \citep{Smith2012,Engler2018} with strong and unexpected traces of both SiO and CO \citep{Lisse2009,Schneiderman2021}. Large amounts of oxygen have also been found at a larger distance \citep{riviere2012}. The main explanation for the presence of gas is a high-speed collision between planetesimals, together with the presence of evaporating icy bodies liberating a mixture of CO, CO$_2$, and H$_2$O out of which oxygen is produced. 

\begin{table}[hbt]
    \centering
    \caption{Main properties of HD\,172555.}
    \label{tab:star}
    \begin{tabular}{lccc}
        Parameter & Unit & Value  & References \\
        \hline
        RA & hh:mm:ss & 18:45:26.9019 & 1 \\
        DEC & dd:mm:ss & -64:52:16.5417 & 1\\
        $m_V$ & mag & 4.77 & 2 \\
        $B$-$V$ & mag  & 0.20 & 2 \\
        $d$ & pc & 28.8 & 1 \\
        $T_\text{eff}$ & K & 7800 & 3,4\\
        $\log g$ & dex & 4.3 & 3,4\\
        $[$Fe/H$]$ & dex & 0.07 & 3  \\
        Sp. type & & A7V & 5 \\
        Age & Myr & 15-25 & 6,7,8,9,10 \\
        $M_\star$ & M$_\odot$ & 1.7 & 4\\
        $R_\star$ & R$_\odot$ & 1.55 & 4, 11\tablefootmark{$\dagger$}\\
        $L_\star$ & L$_\odot$ & 7.8 & 4\\
        \hline
     \end{tabular}
     \tablefoot{(1) \citealt{Gaia2020}; (2) \citealt{Hog2000}; (3) \citealt{Erspamer2003}; (4) \citealt{riviere2012}; (5) \citealt{Gray2006}; (6) \citealt{Barrado1999}; (7) \citealt{Zuckerman2001}; (8) \citealt{Binks2014}; (9) \citealt{Mamajek2014}; (10) \citealt{Miret2018}; (11) this work\tablefootmark{$\dagger$}. \\
     \tablefoottext{$\dagger$}{We used a modified version of the infrared flux method (IRFM; \citealt{Blackwell1977,Schanche2020}), measuring a stellar radius of $R_\star$=1.550$\pm$0.012\,$R_\odot$. It agrees with the $R_\star$=1.56\,R$_\odot$ derived in \citet{riviere2012}. See \citet{Bonfanti2021} and \citet{Wilson2022} for details and application of the IRFM.}
     }
\end{table}

Exocomets were discovered in the HD\,172555 system through $\beta$\,Pic-like varying spectroscopic signatures of cometary transits in the Ca\,II doublet of the star in the optical band observed at high resolution (\citealt{kiefer2014a}; K14, hereafter). Later, they were also observed in the UV in the CII-CII$^\star$ lines \citep{Grady2018}. The transiting clouds are made of gas evaporated from inner nuclei on eccentric orbits that cross the star's line of sight with certain transverse speeds and radial velocities \citep{Beust1990}. The main observed events derived from an exocomet transit and already witnessed in a few young systems have the following properties: i) they are variable, possibly deep and Doppler-shifted, spectroscopic absorptions features due to an extended cloud of sublimed ions covering the solid angle of the stellar surface  \citep{Ferlet1987, Beust1990, Lagrange1992, Vidal1994, Kiefer2014b}; and ii) they experience photometric transits due to the passage of the head and tail of the dust coma in front of the star \citep{Lecavelier1999, Kiefer2017, Rappaport2018, Zieba2019, lecavelier2022}. 

Ions that have been observed using the technique of transit spectroscopy place indirect constraints on the dust from which they have sublimated and further ionised \citep{Beust1993,Kiefer2014b}. Photometry has the advantage over spectroscopy in being able to achieve a direct characterisation of the dust evaporating from the cometary nuclei~\citep{Lecavelier1999}. The detection of 30 exocomet transits in the Transiting Exocomet Survey Satellite (TESS) photometric light curves of $\beta$\,Pic allowed for the first statistical comparison of the exocomets' population in a young systems with the Solar System comets~\citep{lecavelier2022}. It showed that the inner nuclei of Jupiter family comets (JFC, hereafter) and $\beta$\,Pic's exocomets follow a similar radius distribution with radii 1--10\,km and a $dN$$\propto$$R^{-3.6\pm0.8}$\,\!$dR$. Collisional relaxation in a population of small bodies can produce such a distribution.
Our objective is to directly observe the dust produced by the HD\,172555 exocomets, and enable the characterisation of the small bodies in this young system.

In this paper, we report on high-precision photometric monitoring of HD\,172555 with the CHaracterising ExOPlanets Satellite (CHEOPS; \citealt{Benz2021}). In Section~\ref{sec:cheops} we summarise the instrumental configuration and the CHEOPS data reduction. In Section~\ref{sec:delta_scuti} we analyse the $\delta$\,Scuti variations observed in HD\,172555's CHEOPS light curve. In Section~\ref{sec:transient} we report the search for transients in the light curve detrended from the identified $\delta$\,Scuti oscillations. We compare our result to the K14 exocomets in Section~\ref{sec:discussion}. We give our conclusions in Section~\ref{sec:conclusion}.

\section{CHEOPS observations}
\label{sec:cheops}
\subsection{Observations' settings and data reduction}

HD\,172555 was observed within the frame of the CHEOPS Dusty Debris Disk Guaranteed Time Observation (DDD-GTO) program. It was observed for 30 consecutive CHEOPS orbits ($\sim$1.645\,h each), for a total duration of 2.1\,days. Table~\ref{tab:log_obs} gathers details about these CHEOPS observations. The exposure time was set to 0.7 s, leading to a cadence, including readout time, of $\sim$ 43.2 s. The sub-array frames were automatically processed with the CHEOPS Data Reduction Pipeline (DRP; \citealt{hoyer2020}), including smearing, cosmic ray hits, background and stray-light corrections, and finally default aperture photometry extraction. 

\begin{figure}[hbt]
\centering
\includegraphics[width=89.3mm]{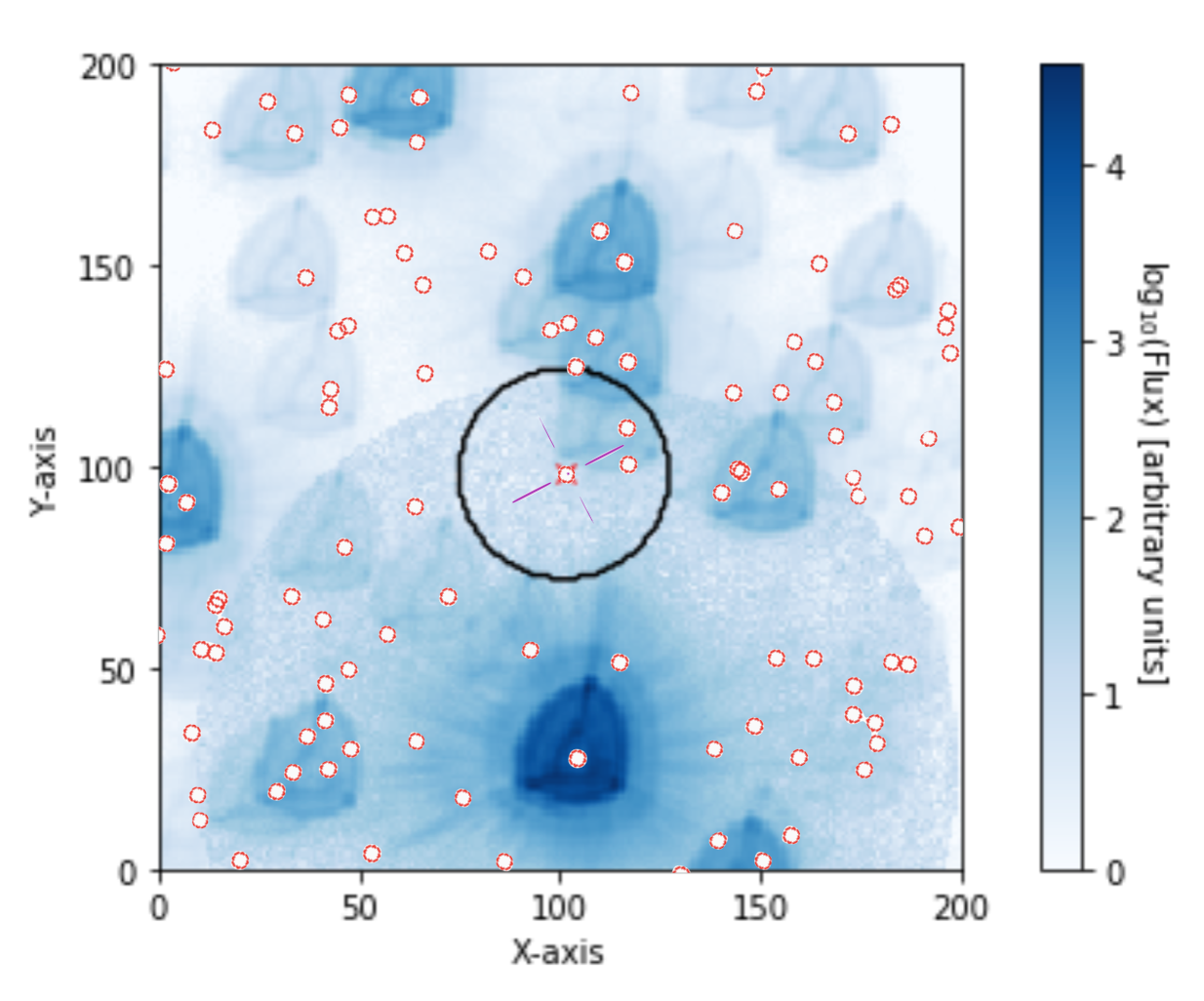}
\caption{Field of view around HD\,172555 showing only background stars with their CHEOPS triangular PSF. East is downward and north is leftward. Gaia identified objects are indicated as red circles and the large black circle represents the optimal aperture of CHEOPS with a size of 25\arcsec. Apart from HD\,172555, the three closest objects within the aperture have G magnitudes of 18.5, 15, and 20 in order of increasing distance from the centre of the aperture. The brightest object in the east is the M-dwarf companion CD-64\,1208 with G = 8.9\,mag.}
    \label{fig:aperture}
\end{figure}

\begin{figure}[hbt]
    \centering
    \includegraphics[width=89.3mm, clip=true]{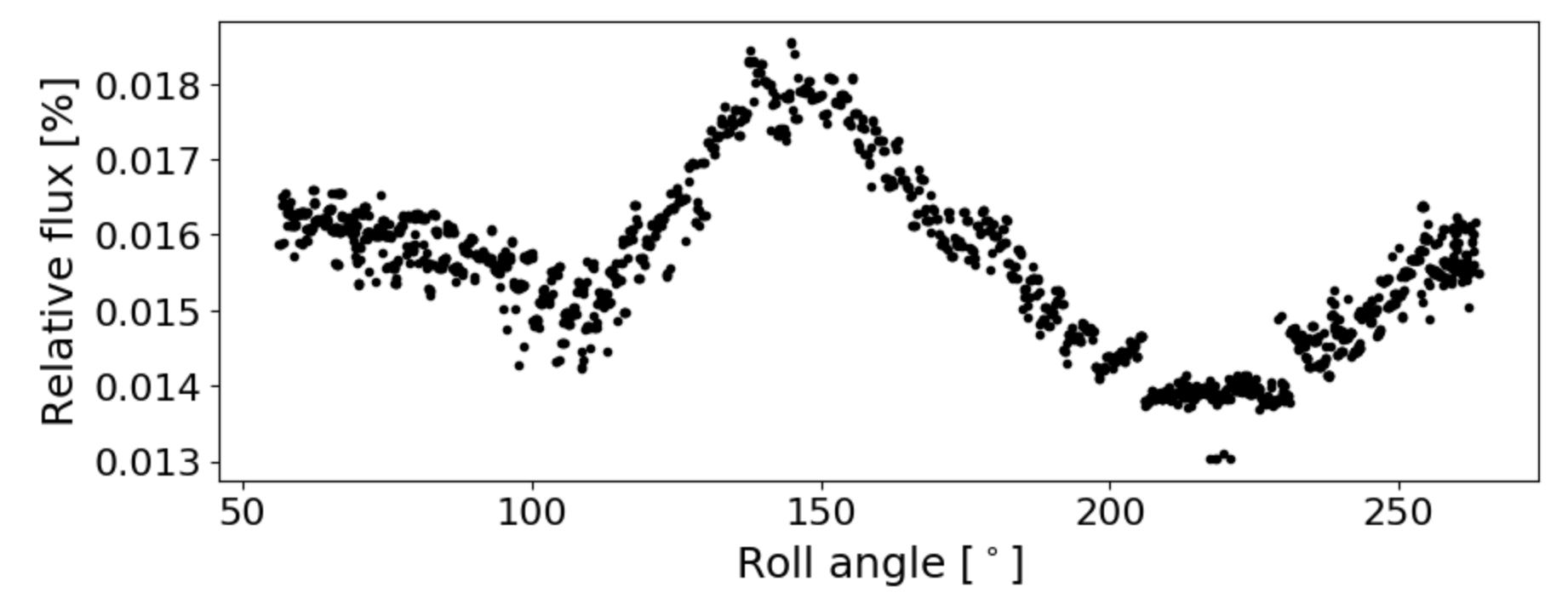}
    \caption{Flux of the simulated contaminant stars in the aperture (black points) as a function of the roll angle (bottom panel) relative to the flux of the target (in percent).}
    \label{fig:rollangle}
\end{figure}

The CHEOPS point spread function (PSF) is relatively extended because of the defocus design of the instrument. The default aperture is defined with a radius of 25 pixels (or equivalently 25\arcsec). The spacecraft rolls with a period of 1.645\,h. While it rolls, background objects enter and leave the extended photometric aperture, leading to periodic contamination effects. The closest neighbours which PSF may enter the default aperture around HD172555 have a magnitude difference of 4.2\,mag with HD\,172555 (see Fig.~\ref{fig:aperture}). The DRP provides an estimate of the contamination effects due to background objects. The average contamination level that is found for HD\,172555's light curve is 155\,ppm, which is on the same order of magnitude as the combined differential photometric precision (CDPP; \citealt{hoyer2020}) of the CHEOPS flux at 10\,min cadence, or $\sim$180\,ppm. However, the full amplitude of the correlation between the roll angle and contamination is 50\,ppm (Fig.~\ref{fig:rollangle}), thus it is 3.6 times smaller than the CDPP. 
 
 \begin{table}[hbt]
     \centering
     \caption{Log of CHEOPS observations}
     \label{tab:log_obs}
     \begin{tabular}{lcc}
         Parameter & Unit & Value \\
         \hline
        DRP Version &  & 13.1.0 \\
        Program ID & & CHEOPS-10 \\
        Program PI & & Gyula Szab\'o \\
        Obs. ID & & 1514950 \\
        CHEOPS orbit & hours & 1.645\\
        Aperture radius & \arcsec & 25 \\
        Source magnitude & CHEOPS band & 4.7 \\
        RA & hh:mm:ss & 18:45:26.9019  \\
        DEC & dd:mm:ss & -64:52:16.5417 \\
        Date start & MJD &  59\,381.445 \\
        Date end & MJD &  59\,383.501 \\
        Total duration & days & 2.06 \\
                       & hours & 49.3 \\
        Exposure time & s & 0.7\\
        Exposures stacking order &  & 24\\
        Total integration time & s & 16.8  \\
        Number of frames &  & 2272 \\
        Number of flagged frames & & 122 \\
        point-to-point RMS & ppm & 1581 \\
        CDPP(10 min) & ppm & 180 \\
        CDPP(3 h) & ppm & 101 \\
        \hline
     \end{tabular}
 \end{table}
 
\subsection{The HD\,172555 light curve}
\label{sec:lightcurve}
\begin{figure*}
    \centering
    \includegraphics[width=198.6mm]{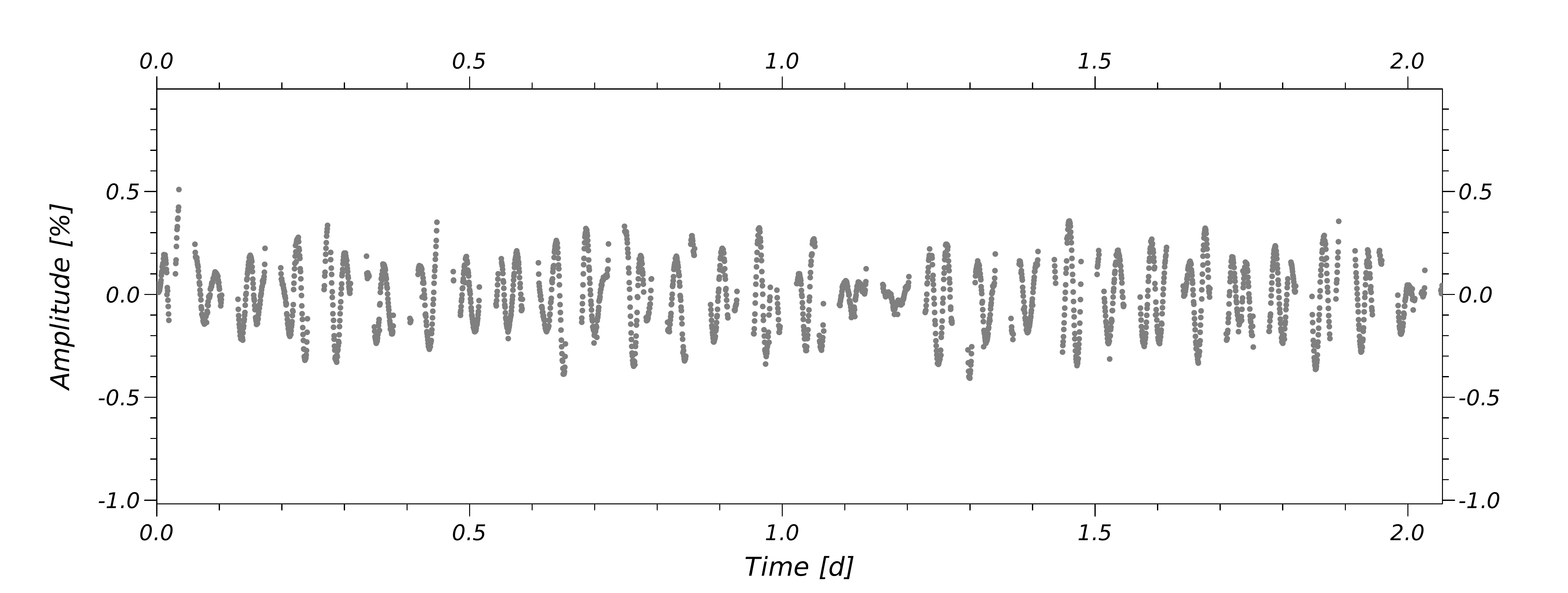}
    \caption{CHEOPS light curve of HD\,172555, with $t_0$=0 at MJD-59381.445.}
    \label{fig:LC}
\end{figure*}

The CHEOPS light curve of HD\,172555 is shown in Fig.~\ref{fig:LC}. The observations are not continuous and have a large number of interruptions or gaps, about 45\% of the total monitoring duration. Those are mainly due to the passage of the star behind the Earth, and crossings of the spacecraft over the South Atlantic Anomaly. 

As summarised in Table~\ref{tab:log_obs}, the root mean square (RMS) of the relative flux in the light curve is 1581\,ppm, but the CDPP that ignores large-scale variations is much lower, 180\,ppm at a 10\,min cadence and 100\,ppm at 3\,h cadence. This provides a better estimate of the short-cadence flux dispersion. The light curve is indeed strongly dominated by large variations that are of stellar origin. 

\begin{figure*}
    \centering
    \includegraphics[scale=0.5]{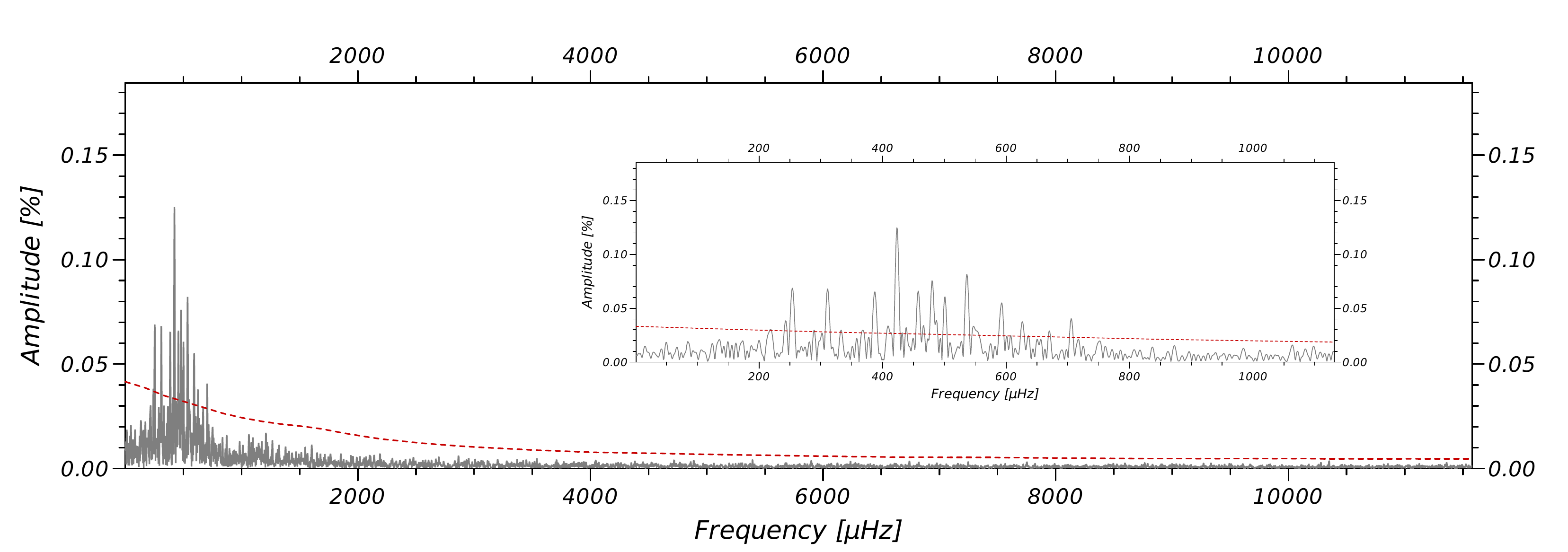}
    \caption{LSP of the CHEOPS light curve of HD\,172555. The red line corresponds to a 4$\sigma$ significance. The Nyquist frequency is at 11\,579 $\mu$Hz. The insert shows a zoomed window on the low frequency domain $<$1,100 $\mu$Hz.}
    \label{fig:LC_LSP}
\end{figure*}

  The time scales of the variability can be clearly seen in the Lomb-Scargle periodogram (LSP), where the variations dominate in the range $\sim$200--800\,$\mu$Hz (Fig.~\ref{fig:LC_LSP}). We attributed these strong periodic variations to stellar oscillations, which, given the position of the star ($T_\text{eff}$=7800\,K, $\log g$=4.3; Table~\ref{tab:star}) in a $T_{\rm eff}-\log g$ diagram such as Fig. 10 of~\citet{Uytterhoeven2011} and the oscillation frequencies that have been identified, are of a $\delta$\,Scuti nature. 
  
  We aim to search for weaker signals, such as transits, that are possibly present in the light curve. At this stage of the reduction, those would still be hidden behind the strong oscillations. Our purpose now is thus to remove this variability. 

\section{HD\,172555's $\delta$\,Scuti oscillations}
\label{sec:delta_scuti}

First, we stress that from the frequency extraction carried out below, there is no intention to use the stellar oscillations to attempt to model the star with asteroseismology. The observation duration of about $T = 2.1$\,days is too short for this purpose, as it results in a resolution of only $1/T = 5.6\,\mu$Hz. Instead, our aim is to clean the light curve as much as possible, from any periodic variability of stellar or instrumental origin, in order to search for transient signatures.

To remove the periodic oscillations, we applied the standard pre-whitening technique \citep{Deeming-75} using the tool Felix \citep[Frequency Extraction for LIght curve eXploration;][]{Charpinet2010,Zong2016}. In short, in the LSP, we identified the frequency and amplitude of the highest-amplitude peak, which are used as initial guesses in a subsequent non-linear least square (NLLS) fit of a cosine wave in a time domain using the Levenberg-Marquardt algorithm. The fitted wave of derived frequency, amplitude, and phase was then subtracted from the light curve. The operation was repeated as long as there was a peak above a pre-defined threshold, defined as a given level of the signal-to-noise ratio (S/N). The S/N=1 level -- the noise -- is defined locally as the median of the points within a gliding window (centred on each point of the LSP) of $\sim$300 times the resolution of the data. This median was re-evaluated at each step of the pre-whitening, that is each time a peak was removed. 

\begin{figure}
\includegraphics[width=89.3mm]{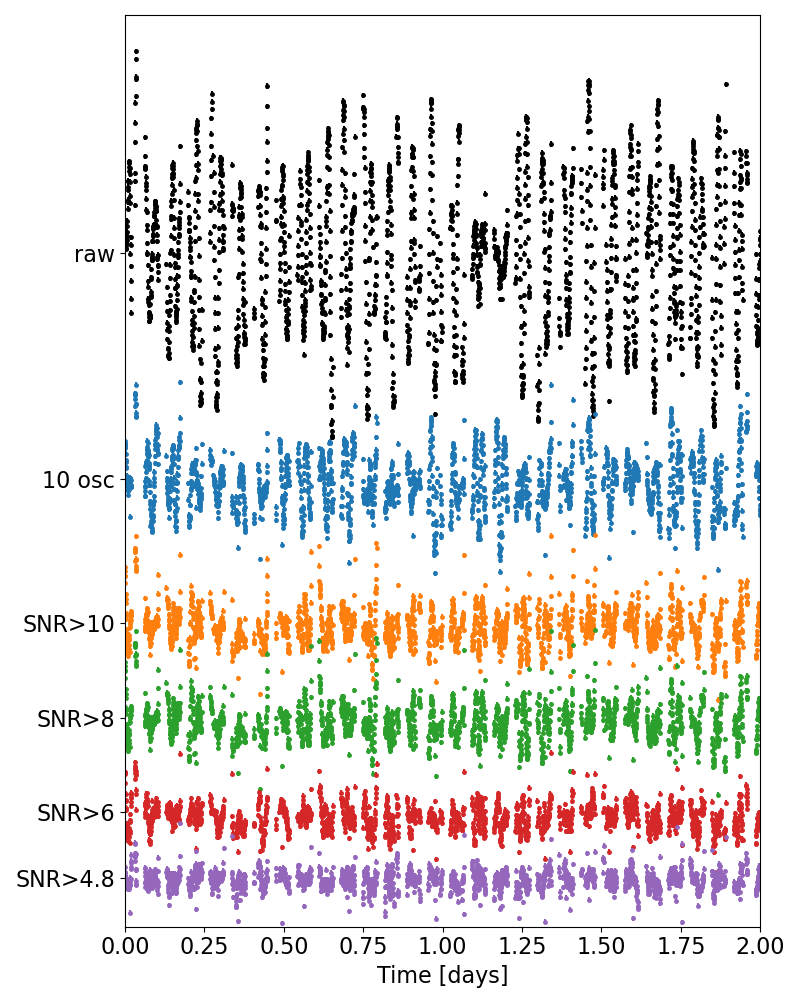}
\caption{\label{fig:residuals_compared} HD\,172555's light curve after the removal of $\delta$\,Scuti variations with different degrees of precision (see text). From top to bottom: Raw light curve (black), the ten dominant frequencies subtracted (blue), and the oscillations subtracted with S/N>10 (orange), S/N>8 (green), S/N>6 (red), and S/N>4.8 (purple). The vertical scale is common to all curves, allowing for the effect of the different $\delta$\,Scuti removal schemes on the residuals' amplitude to be compared.}
\end{figure}

We detrended the light curve from the oscillations at different levels of significance (as determined by the S/N), from relatively aggressive to conservative. The minimum significance required for identifying a peak to be removed is 4$\sigma$, that is a false alarm probability of $3.2\times10^{-5}$. This 4$\sigma$ significance level can be converted into an S/N=$x$ level. To determine the value of $x$, we used the approach developed in \citet{Zong2016}: using the same time sampling (i.e. a cadence of 43.2s) and the same window (i.e. accounting for gaps) as in the real light curve, we simulated 10\,000 pure Gaussian white-noise light curves. For a given S/N threshold, we then searched for the number of times that at least one peak in the LSP of these artificial light curves (that are, by construction, just noise) happen to be above this threshold. We obtained the false alarm probability by dividing by the number of tests (10,000 here). We found that the threshold corresponding to a 4$\sigma$ significance (false alarm probability of $3.2\times10^{-5}$) is S/N=4.8. Table~\ref{app:Freq} presents the properties of the periodic variations extracted in the CHEOPS light curve down to S/N$=$4.8. %Removal of these frequencies have been done down to various thresholds in order to clean the light curve for detecting transients (see below).
Frequencies are detected from 103.4\,$\mu$Hz almost up to the Nyquist frequency of 11\,579\,$\mu$Hz. From $f_{34}$ (1011.23\,$\mu$Hz) and beyond, it should be noticed that all peaks correspond to multiples of the CHEOPS orbital frequency (orbital period of 98.9\,min). Thus, contaminations correlated to the roll angle of the spacecraft with amplitudes $<$200\,ppm were still present in the light curve, which FELIX succeeded to identify and remove. More conservative thresholds that we applied for removing periodic variabilities are S/N$>$ 6, 8, \& 10, and finally, most conservatively, removing only the ten dominant frequencies. The Figure~\ref{fig:residuals_compared} shows the effect of the different detrending schemes. 

\section{Transient signature search}
\label{sec:transient}
\subsection{Hint of a transit-like event in the binned residual light curves}

 Our purpose is now to analyse the detrended light curves of HD\,172555 in the search for any transient events of circumstellar origin. We expect to witness them in such a young inclined system. 
Some short-scale ($<$1\,h) variabilities are still present in the detrended light curves, which cannot be of circumstellar origin because a central transit of the star at a distance $>$1\,$R_\star$ is at least 1.2\,h long. We thus averaged out this variability by applying a mean filter to the data, that is equivalent to binning. A window size of 1.645\,h, equal to the CHEOPS orbital period, offers a good compromise between decreasing noise, removing excess variability, synchronising the bins and the gaps, and enhancing the prominence of $>$2\,h variations.

Before calculating the binned light curve, we linearly interpolated the data on a grid of epochs with a constant step of 43\,s. Epochs within gaps were treated as missing data. For each bin, the binned flux is the mean among all epochs within the binning window. The equation that we used to calculate error bars is:

\begin{equation}
    \sigma^\text{bin}_i = \sqrt{\sum_{j=1}^N \frac{\sigma_{i,j}^2}{N^2} + \frac{s_i^2}{N} }
,\end{equation}

where $i$ is the index of a binned point, $j$ is a sub-index of the binning window going from 1 to $N$, $\sigma_{i,j}$ is the intrinsic (CHEOPS) flux error bar of the considered epoch in the light curve, and $s_i$ is the standard deviation of flux within bin $i$. The leftmost term is the quadratic sum of the flux errors of all epochs within bin $i$ divided by $N^2$; this is the variance on the mean of a set of random variables. Missing data ($\sim$45\% of the time span) are taken into account by attributing an uncertainty to them that is equal to one standard deviation of the full light curve $\sigma_\text{lc}$. The rightmost term is the square of the standard deviation of flux within bin $i$, divided by $N$; this is the estimator of the variance of the mean within a sample of values (here bin $i$).

\begin{figure}
    \centering
    \includegraphics[width=89.3mm, clip=true, trim=0 70pt 0 130pt]{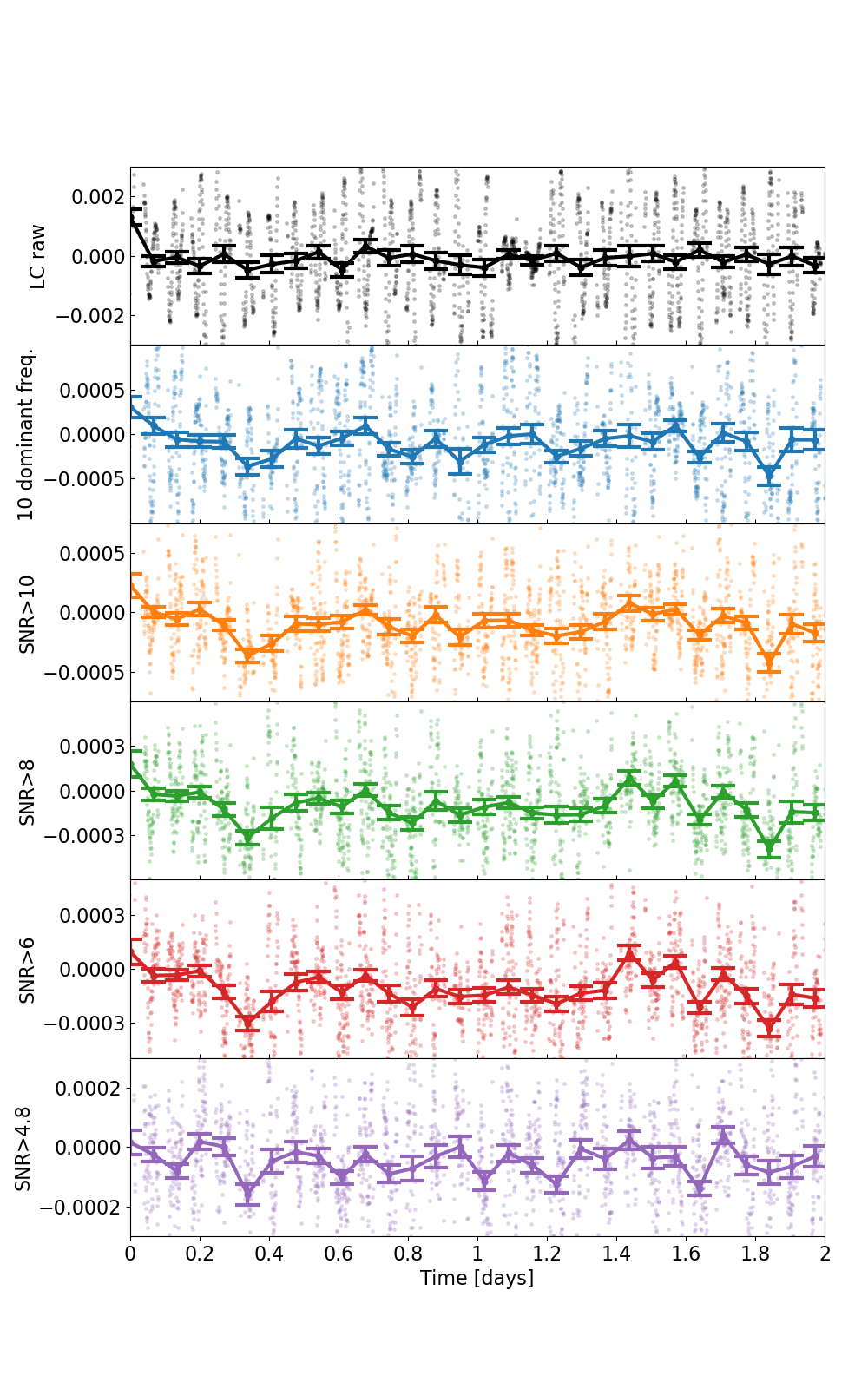}
    \caption{Binning of the raw and residual light curves with a 1.645\,h bin. The colour code is the same as in Fig.~\ref{fig:residuals_compared}. The dots show, in comparison, the original, non-binned, light curve.}
    \label{fig:binning_compared}
\end{figure}

Figure~\ref{fig:binning_compared} shows the effect of binning on the different detrended light curves obtained after subtracting $\delta$\,Scuti oscillations and other periodic variations at various thresholds (see also Section~\ref{sec:delta_scuti}). In the binned detrended light curves, there is one prominent feature, an absorption at about $t$=0.3\,days with an apparent depth 300--500\,ppm, common to every detrending scheme. Another series of dips attract attention near the edge within 1.6--1.8\,days. However, they are less prominent than the t=0.3-days feature, and with a short duration, they are similar to the bin period (1.645\,h), and thus more likely to originate from noise. For this reason, in this work, we only focus on the t=0.3-days feature, our transient candidate.

In all the detrended light curves, residual $\delta$\,Scuti oscillations dominate over other stochastic noise. Even though it is apparently not periodic in the 2-days long data, we cannot fully exclude that such a transient signal is due to the constructive interference of several $\delta$\,Scuti oscillations. We have checked, however, that the frequencies involved in a transient with such a duration ($\sim$0.4\,days) rather stand below 100\,$\mu$Hz (or $T$=0.1\,days). As shown in the light curve LSP (Fig.~\ref{fig:LC_LSP}), a signal at these frequencies with an amplitude of $>$300\,ppm is well separated from the majority of frequencies (500--800\,$\mu$Hz) attributed to the $\delta$\,Scuti pulsator. This suggests that the identified transient is not an artefact of the stellar oscillations.

Furthermore, in Section~\ref{sec:systematics}, we are able to exclude that this transient is due to instrumental or reduction systematics. In Section~\ref{sec:transient_fit}, assuming this transient might be produced by the passage of an exocomet in front of HD\,172555, we fitted it with an exocomet transit model with a good match. We then show in Section~\ref{sec:injection} that an injected exocomet transit of a few hundred parts-per-million is expected to survive the $\delta$\,Scuti oscillations' removal process applied above, down to S/N$\sim$6. 

\subsection{Systematics}
\label{sec:systematics}
\begin{figure}
    \centering
    \includegraphics[width=89.3mm,clip=true,trim=0 80pt 0 0]{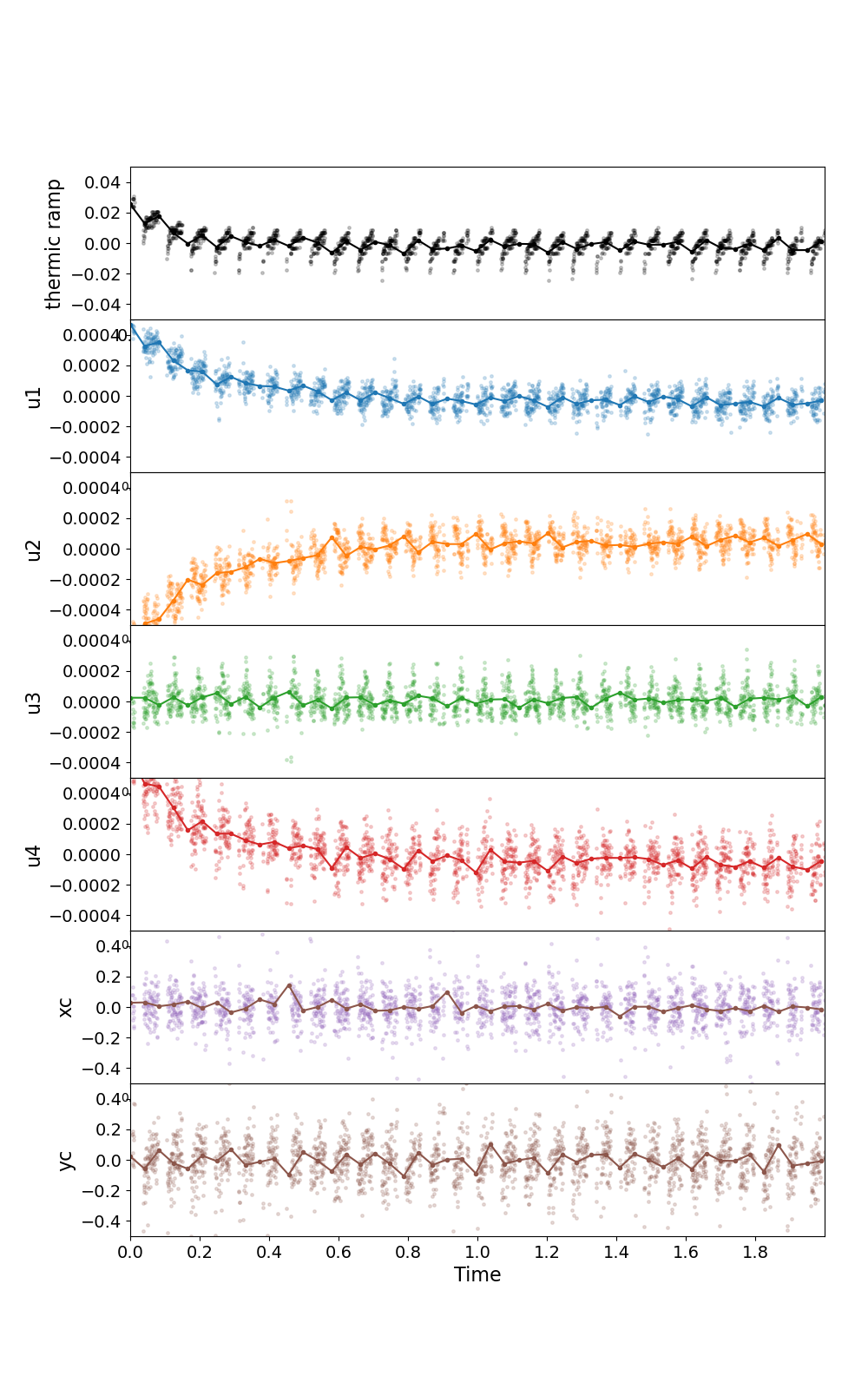}
    \caption{Systematics variations, from top to bottom: the recording from one of the thermal front sensors (in arbitrary units); the PSF principal components' coefficients $u_1$, $u_2$, $u_3$, and $u_4$ (in arbitrary units); and the PSF centroid $(x_c,y_c)$ (in pixels).}
    \label{fig:systematics}
\end{figure}

We first compared the detrended light curve with the systematics of CHEOPS: thermal ramp, PSF principal component coefficients, and centroid variations (Fig.~\ref{fig:systematics}). The variations were binned, as the light curves above, with a 1\,h timestamp. We used the reduction pipeline called PIPE to extract the PSF photometry (Brandeker et al. in prep.; see also descriptions in \citealt{szabo2021,Brandeker2022}), including the PSF principal component coefficients and the centroid variations. 

We identified a long-term variation with an exponential-decay shape. This variation is indeed present in the data as well and commonly identified as being due to thermal relaxation of the telescope tube, as recorded by the thermal front sensor of CHEOPS \citep{deline2022}. However, we did not observe short timescale transient-like signals in these diagnostics. This suggests that the transient signal is of astrophysical origin, and not of instrumental origin. 

Moreover, the PIPE PSF photometry gives an independent photometric extraction with potentially different systematics from the aperture photometry of the DRP. In the present case, PIPE gives very similar results to the DRP with only marginal improvements in the mean absolute deviation (163\,ppm for PIPE compared to 168\,ppm for the DRP). The transient signal is still present. In the rest of the analysis, we use only the DRP reduction.

\subsection{Transit signature fit}
\label{sec:transient_fit}
HD\,172555 is known to be surrounded by a debris disk and transiting exocomets (K14; \citealt{Grady2018}). Given this information, we make the hypothesis that the identified transient absorption feature is due to the transit of an exocomet in front of HD\,172555. Such exocomet's photometric transit has already been witnessed in $\beta$\,Pictoris \citep{Zieba2019,Pavlenko2022,lecavelier2022}. 

We could fit this feature with a photometric exocomet transit model of the relative flux variation $\Delta F/F$. We followed \citet{lecavelier2022} and used a 1D empirical model of an exocomet transit based on thorough numerical simulations of transit shapes of an evaporating comet by \citet{Lecavelier1999}: 
\begin{equation}\label{eq:model}
    \frac{\Delta F(t)}{F(t)} = K \left(e^{-\beta (t-t_0)\,H(t-t_0)} - e^{-\beta (t-t_0-\Delta t)\, H(t-t_0-\Delta t)} \right)
,\end{equation}

with $H(x)$ being the Heaviside function. In this model with four parameters, $t_0$ is the time of the beginning of transit of the head of the exocomet; $\beta$ is the speed of the transit of one scale length of the cometary tail; $\Delta t$ is the duration of the transit of the comet nucleus; and $K$ is a scalar amplitude related to absorption depth $AD$ -- that is, the deepest relative flux variation during the transit  -- through $AD$=$K\,(1-\exp(-\beta \Delta t))$. We used $\beta$ from 0 to 100\,days$^{-1}$ and $\Delta$t from 0.04 to 0.4 days, corresponding to periastron distances ranging from about 0.01 to 0.13 au.

More importantly, from these quantities, we could estimate physical properties of the exocomet nucleus using a scaling relation based on the well-constrained properties of the bright solar system's Hale-Bopp comet~\citep{Jewitt1999}. It has a radius of $\sim$30\,km with a dust production rate of 2$\times$10$^6$\,kg\,s$^{-1}$ at 1\,au that can be translated into the following scaling relation \citep{lecavelier2022}:
\begin{equation}\label{eq:halebopp}
\dot M_\text{dust, 1\,au} = 2\times 10^6\,\text{kg}\,\text{s}^{-1}
\left(\frac{r_\text{nucl}}{30\,\text{km}}\right)^2 \left(\frac{L_\star}{L_\odot}\right)
,\end{equation} 
with $\dot M_\text{dust, 1\,au}$ being the dust-mass production rate of the exocomet if it were at 1\,au from the central star, $r_\text{nucl}$ being the radius of the solid comet nucleus, and $q$ being the periastron distance to the central star. The Hale-Bopp dust production rates, in agreement with \citet{Lecavelier1999}'s model assumptions, follow a $\dot M$$\propto$$1/q^2$ law, with a decrease by a factor $\sim$\,\!$10$ when distance increases from 1 to 3\,au~\citep{Jewitt1999}. 

We also used equations relating the model parameters to physical properties of the exocomets. They are based on the following identities\,\citep{lecavelier2022}:

\begin{align}
    & AD = 5\times 10^{-5} \left(\frac{\dot M_\text{1\,au}}{10^{5}\,\text{kg}\,\text{s}^{-1}} \right) \left(\frac{q}{1\,\text{au}} \right)^{-1/2} \left(\frac{M_\star}{M_\odot} \right) \\
    & \Delta t_\text{transit} =  \frac{\bar L_\text{path}}{v_\text{transit}} \\
    &\text{with}  \quad v_\text{transit} \approx \sqrt{\frac{2\,G M_\star}{q}} 
     \quad \text{and} \quad \bar L_\text{path} = \frac{\pi R_\star}{2}. \nonumber
\end{align}
The duration of transit $\Delta t_\text{transit}$ is given by the average path length through the stellar disk divided by the transit velocity $v_\text{transit}$. Using the stellar parameters of HD\,172555 given in Table~\ref{tab:star}, these identities translate into the following:

\begin{align}
    & q \approx 0.013\, \left(\frac{\Delta t}{1\,\text{h}}\right)^2 ~~ \text{au} \label{eq:lecavmodel1}\\
    & \dot M_\text{1\,au} \approx 3.7\times 10^5 \, \left(\frac{AD}{1000\,\text{ppm}}\right) \sqrt{\frac{q}{0.1\,\text{au}}}  ~~\text{kg\,s}^{-1} \label{eq:lecavmodel2}\\
    & r_\text{nucl} \approx 2.4 \, \left(\frac{\dot M_\text{1\,au}}{10^5\,\text{kg\,s}^{-1}}\right)^{1/2} ~~\text{km.}
    \label{eq:lecavmodel3}
\end{align}

Previous exocomet detections in Ca\,II spectra of the HD\,172555 system have a radial velocity within $\pm$20\,km\,s$^{-1}$ from the star's systemic velocity, and a distance to the star $\lesssim$0.3\,au such that Calcium can sublimate at a rate enabling the ion cloud to cover a significant portion of the line of sight \citep{Beust1990}. Their transit should thus happen shortly before or after the periastron passage along their near-parabolic orbit. In the above formula, we thus assumed that the exocomet is transiting the stellar disk at a locus close to the periastron, with a distance to the central star $\sim$$q$.

\begin{figure}
    \centering
    \includegraphics[width=89.3mm,clip=true,trim=0 220pt 0 0]{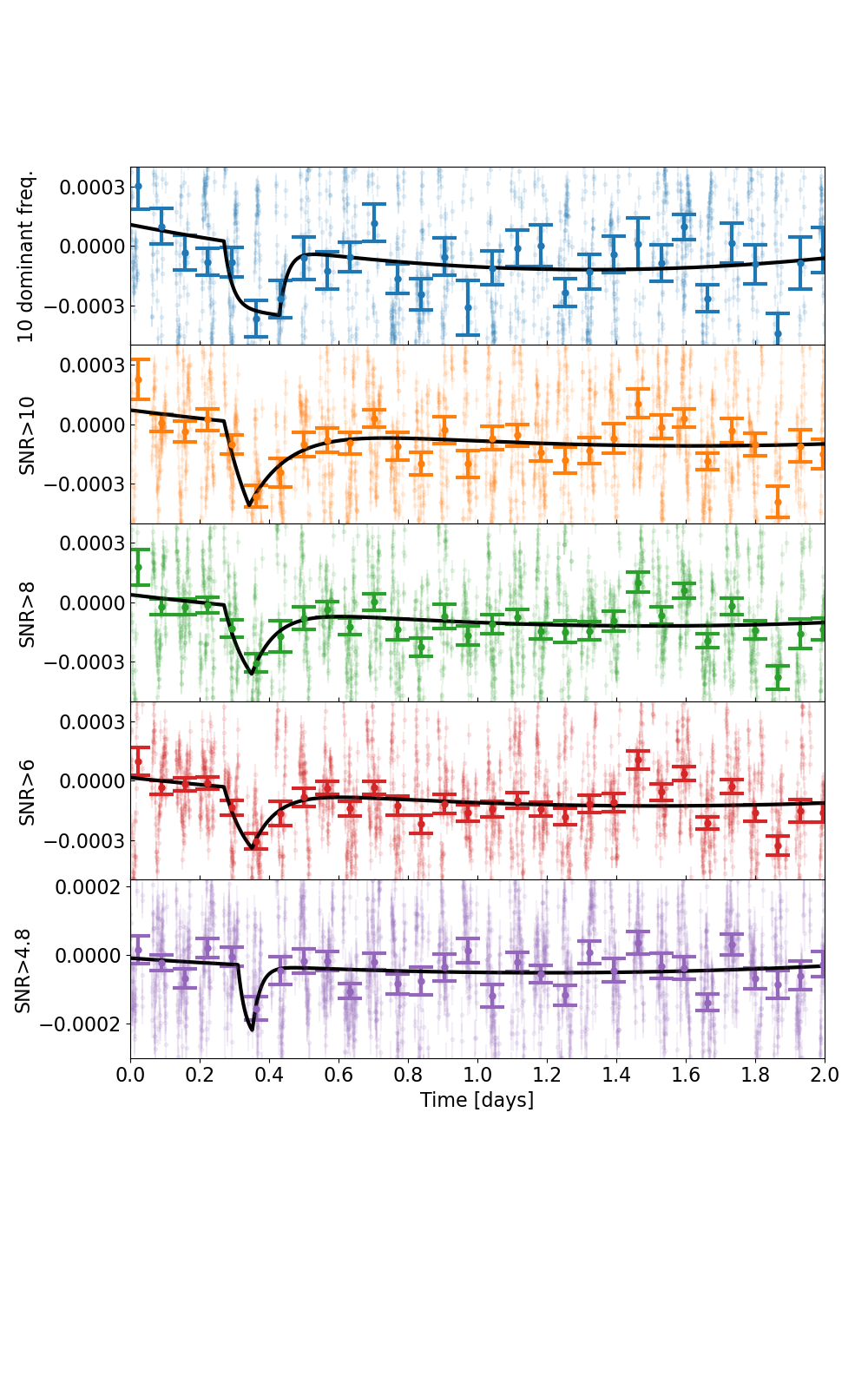}
    \caption{Transient signature detected at $t-t_0$=0.3\,day and modelled by a transiting exocomet with the model given in Eqs.~\ref{eq:model}-\ref{eq:lecavmodel3}, compared to the 1-hour binned detrended light curves. From top to bottom, the colour codes identify the same datasets as in Fig.~\ref{fig:binning_compared}.}
    \label{fig:model_fit}
\end{figure}

We fitted all differently detrended light curves with this exocomet transit model and a 2-degree polynomial to account for the long-term trend identified as a thermal relaxation. The results of the fit are summarised in Table~\ref{tab:result} and the model compared to the binned light curves is shown in Fig.~\ref{fig:model_fit}. We derived the $\chi^2$ and an $F$ test, comparing the cases of fitting an exocomet transit profile plus a 2-degree polynomial, and as the null hypothesis, a 2-degree polynomial only.

The depth of the signature indeed decreases for detrending deeper than S/N$>$8. The shape of the transit profile also seems strongly affected by residual $\delta$\,Scuti oscillations if only the ten\,dominant frequencies are subtracted. It can be seen in the null hypothesis $\chi^2_\text{null}$ that oscillations still dominate the error budget of the detrended light curve, even if they are removed down to S/N=4.8. Nevertheless, the adjunction of the transit signature to the fitted model always significantly improves the $\chi^2$, with an F test $>$6.8 (p-value$<$10$^{-5}$).

It is unclear, however, at what S/N threshold of the $\delta$\,Scuti detrending schemes would the derived parameters be the closest to the true transiting exocomet, if any. To understand how $\delta$\,Scuti removal may also remove transit signal, we have to test transit injection recovery.

\begin{table*}[]
    \centering
    \caption{Fit results with the exocomet transit model of Eqs.~\ref{eq:model}-\ref{eq:lecavmodel3}.}
    \label{tab:result}
    \begin{tabular}{lccccccc}
        Parameters & Unit &  10 dom. freq. &  S/N$>$10 &  S/N$>$8 & S/N$>$6 & S/N$>$4.8  \\
        \hline
        $t_0$ & BJD-$t_\text{min}$ & 0.273$\pm$0.015 & 0.271$\pm$0.012 & 0.271$\pm$0.011 & 0.270$\pm$0.010  & 0.310$\pm$0.020                      \\
        $\beta$ & day$^{-1}$ & 43$\pm$38 &  8.6$\pm$2.3 & 14.3$\pm$4.5 & 14.3$\pm$4.3  & 44.0$\pm$29.4 \\
        $\Delta t$ & hour & 3.75$\pm$0.40 &  1.69$\pm$0.59 & 1.90$\pm$0.54  & 1.93$\pm$0.50 & 1.00$\pm$0.57  \\
        $K$ & ppm & 337$\pm$73 & 920$\pm$380 & 497$\pm$187 & 435$\pm$148  & 224$\pm$11\\
        \hline
        $\chi^2$ & DoF=2212 & 284776 & 119804 & 78834 & 56445 & 31623 \\
        $\chi^2_\text{null}$ & DoF=2216 & 289145 & 123559 & 81087 & 58240 & 32009\\
        F-test & & 8.5 & 17.3 & 15.8 & 17.6 & 6.8\\
        $p$-value & & $8.6\times10^{-7}$ & $5.2\times10^{-14}$ & $9.2\times10^{-13}$ & $3.2\times10^{-14}$ & $2.1\times10^{-5}$\\
        \hline
        $AD$ & ppm & 336$\pm$73 & 417$\pm$219 & 336$\pm$148 & 297$\pm$119 & 188$\pm$110 \\
        $q$  & au & 0.186$\pm$0.014 & 0.038$\pm$0.010 & 0.047$\pm$0.010 & 0.049$\pm$0.009 & 0.013$\pm$0.005 \\
        $\dot M_\text{dust, 1\,au}$ & $10^5$\,kg\,s$^{-1}$& 1.81$\pm$0.40 & 1.01$\pm$0.55 & 0.92$\pm$0.41 &  0.82$\pm$0.34 & 0.27$\pm$0.17 \\
        $r_\text{nucl}$ &  km & 3.38$\pm$0.37 & 2.53$\pm$0.68 &  2.40$\pm$0.54 & 2.28$\pm$0.47 & 1.30$\pm$0.41\\
        \hline
    \end{tabular}
\end{table*}

\subsection{Transit injection recovery}
\label{sec:injection}
Here, we check the following: that an exocomet transit signal could indeed survive the $\delta$\,Scuti removal process; at what threshold the transit parameters are best retrieved; and, especially, below what threshold such a transit signal is identified by FELIX and removed as if it were of $\delta$\,Scuti origin.
To do so, we injected a transient into the raw light curve of HD\,172555 prior to the detrending scheme presented in Section~\ref{sec:delta_scuti}. We injected a signal at $t-t_0$=$1.2$\,days similar to the one potentially detected here (Section~\ref{sec:transient_fit}), with a transit depth of 544\,ppm and a full width at half maximum of $\Delta t = 0.25$ days. All parameters of the injected model are given in Table~\ref{tab:injection}. 

\begin{figure}[hbt]
    \centering
    \includegraphics[width=89.3mm]{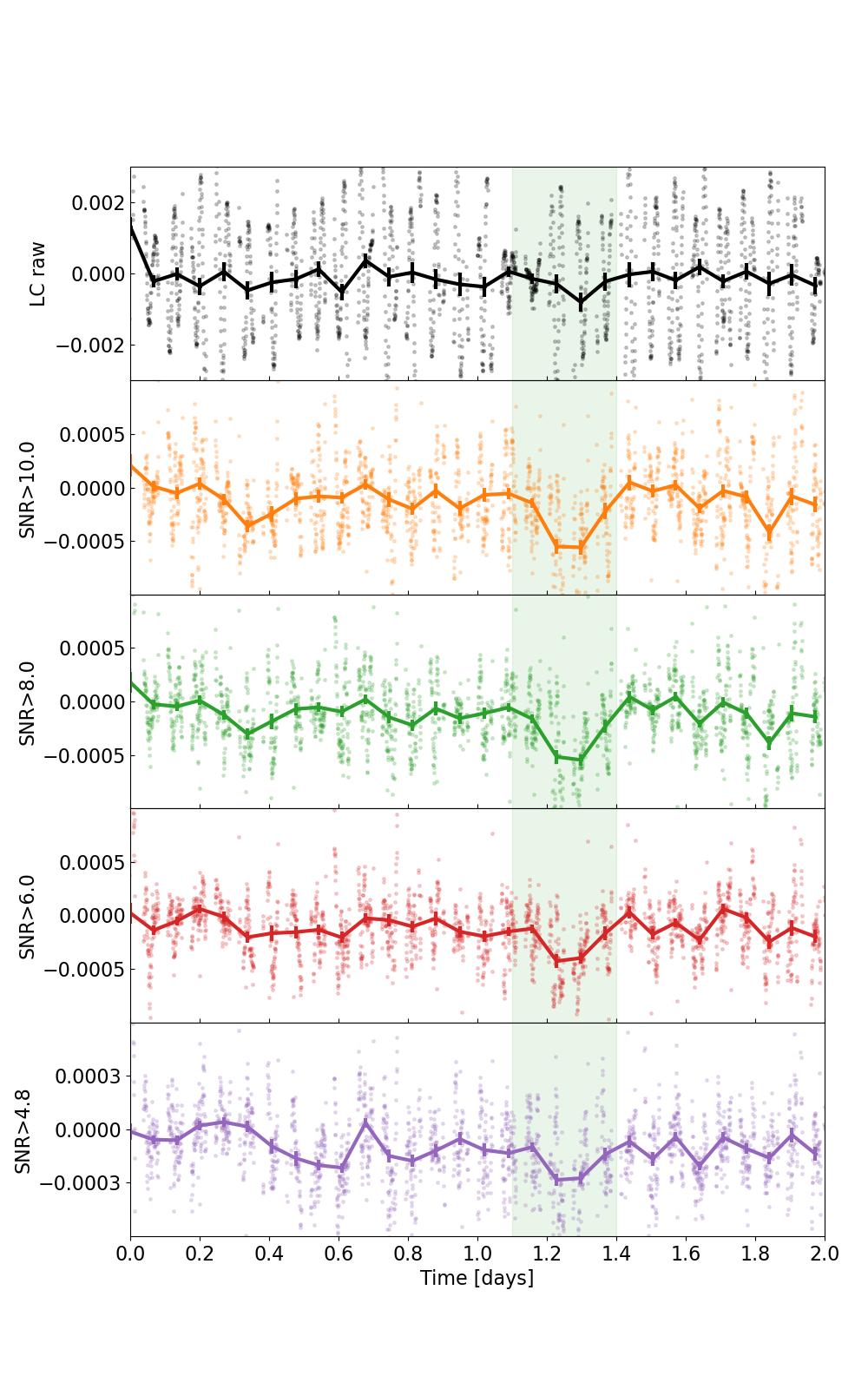}
    \caption{Recovery of an injected transient signal (green filled area) at about $t-t_0$=1.2\,days from detrended light curves where the removed $\delta$\,Scuti oscillations have the same S/N levels as analysed here with the same colour code as in Fig.~\ref{fig:residuals_compared}. The raw light curve with the injected signal is shown on top (black line). The adopted bin step is also 1.645\,h.}
    \label{fig:injection}
\end{figure}

Figure~\ref{fig:injection} shows the detrended light curves when removing the periodic variabilities down to S/N$>$4.8, 6, 8, and 10. The injected signal is clearly detected in all the detrending schemes. For the 'S/N$>$4.8' scheme, the injected signal is strongly affected by the removal of oscillations and the absorption depth is strongly damped to below 544\,ppm (as injected), while the t=0.3-day candidate transient has disappeared. 

In Table~\ref{tab:injection} we added the parameters fitted at the position of the injected transient, showing that the transit time, $\Delta t$, and $K$ parameters derived for the detrended light curves with S/N$>$8-10 are almost not affected by removal of periodic variabilities. The value derived for $\beta$ is strongly discrepant with the injected value, showing the difficulty to have a good estimate as to the real signal. 
The most accurate estimation of periastron distance $q$, dust production rate $\dot M_\text{dust,1\,au}$, and nucleus radius $r_\text{nucl}$ were obtained by the fit of the injected transit signature in the 'S/N$>$8' detrending scheme.

\subsection{Final adopted parameters of the candidate exocomet}
In this scheme, the candidate exocomet transit model detected in this work, fitted in Section~\ref{sec:transient} with the results shown in Table~\ref{tab:result}, has 
$\beta$=$14$$\pm$$5$ day$^{-1}$, $\Delta t$$=$$1.9$$\pm$$0.5$ h, and $K$$=$$500$$\pm$$200$ ppm. This leads to an absorption  
depth $AD$$\sim$$340$$\pm$$150$\,ppm, a periastron distance $q$$\sim$$0.05$$\pm$$0.01$\,au, a dust production rate $\dot M$$\sim$$(0.9$$\pm$$0.4)\times10^5$\,kg\,s$^{-1}$, and a nucleus size of 2.5$\pm$0.5\,km. This is our best estimate of the true profile, assuming the transient is indeed due to an exocomet transit.

\begin{table*}[hbt]
    \centering
    \caption{Injection model and fits (see text for explanations).}
    \label{tab:injection}
    \begin{tabular}{lcccccc}
        Parameters & Unit & Injected & S/N$>$10 & S/N$>$8 & S/N$>$6 &  S/N$>$4.8\\
        \hline
        $t_0$ & BJD-$t_\text{min}$ & 1.2 & 1.226$\pm$0.005 & 1.232$\pm$0.003 & 1.228$\pm$0.004  & 1.250$\pm$0.003                    \\
        $\beta$ & day$^{-1}$ & 15 & 43.0$\pm$13.0 & 98.1$\pm$31.8 & 58.3$\pm$18.0 &  45.9$\pm$12.4\\
        $\Delta t$ & hour & 2.4 & 2.11$\pm$0.15  & 2.18$\pm$0.08 & 2.02$\pm$0.12  & 1.00$\pm$0.18\\
        $K$ & ppm & 700 & 697$\pm$85  & 582$\pm$44 & 460$\pm$51  & 601$\pm$146 \\
        \hline
        $\chi^2$ & DoF=2211 & & 121972 & 90049 & 56594  & 32371 \\
        $\chi^2_\text{null}$ & DoF=2215 & & 133317 & 100737 & 61623 & 34136\\
        F-test & & & 51.4 & 65.6 & 49.1 & 30.1\\
        \hline
        $AD$ & ppm & 544 & 681$\pm$85 & 582$\pm$44 & 456$\pm$51 & 513$\pm$136 \\
        $q$  & au & 0.037 & 0.029$\pm$0.002 & 0.031$\pm$0.001 &  0.026$\pm$0.002 & 0.007$\pm$0.001 \\
        $\dot M_\text{dust, 1\,au}$ &  $10^5$\,kg\,s$^{-1}$& 1.24 &  1.36$\pm$0.18 & 1.20$\pm$0.09 &  0.87$\pm$0.10 & 0.49$\pm$0.14\\
        $r_\text{nucl}$ &  km & 2.82 & 2.96$\pm$0.18 & 2.78$\pm$0.11 & 2.37$\pm$0.13 & 1.77$\pm$0.23 \\
        \hline
    \end{tabular}
\end{table*}

\section{Comparing this candidate exocomet to K14 spectroscopic detections}
\label{sec:discussion}

It is interesting to make a comparison between the present candidate detection and those reported in K14. It can allow us to understand if they belong to similar or different populations. 

Some parameters of their detections are reported in Table~\ref{tab:kiefer2014}. Using the formulas from~\citet{Kiefer2014b}, we could calculate the evaporation efficiency of all those detections, with the evaporation efficiency defined as the log ratio between the energy spent to evaporate gas and dust from the exocomet nucleus and the incoming energy flux. For HD\,1725555, it is related to the dust production rate, with $d$ being the distance to the star in $R_\star$ as

\begin{equation}\label{eq:eff}
    \eta = \log(\dot M_\text{dust} \,  d^2) - 1.86
.\end{equation}

For variable absorption lines in Ca\,II K \& H spectra due to the occultation of the stellar surface by a cloud of atomic gas evaporated from an exocomet, the depth of the variable lines in Ca\,II K \& H are $\alpha(1-e^{-A})$ and $\alpha(1-e^{-A/2})$, respectively, with the oscillation strength ratio $f_H/f_K$=0.5 (K14; \citealt{Kiefer2014b}). We introduced the cloud-to-star surface ratio $\alpha$ and the optical absorption depth $A$. The evaporation efficiency can be determined from $\alpha$ and $A$ (see \citet{Kiefer2014b} for details) as 

\begin{equation}
    \eta = 9.2 + \log\left[\alpha \left(1-e^{-A}-e^{-A/2}\right)\right]
.\end{equation}

Another useful formula from~\citet{Kiefer2014b} allows for the distance to the star to be expressed with respect to the surface ratio $\alpha$ and the dust production rate: 

\begin{equation}
\alpha = 5 \times 10^{-14} \dot M^{4/3} d^{4/3}    
.\end{equation}

Combining this equation with Eq.~\ref{eq:eff} leads to an expression of the distance, in stellar radius, with respect to the evaporation efficiency and the surface ratio

\begin{equation}
    d = 7.7\times 10^{-9} \, 10^\eta\, \alpha^{-3/4}
.\end{equation}

It is then straightforward to derive the dust production rate at 1\,au, $\dot M_\text{dust, 1\,au}$, again assuming that the exocomet locus is close to its orbit periastron, such that $d$$\sim$$q$ is the periastron distance. Further, applying equations~(\ref{eq:lecavmodel1}-\ref{eq:lecavmodel3}) allows one to determine, for each K14 detection, an estimation of transit duration $\Delta t$ and the photometric transit absorption depth. All derived parameters are reported in Table~\ref{tab:kiefer2014}. The error bars were obtained by: i) sampling split-normal distributions of $\alpha$ and $A$ by drawing 10,000 samples; ii) applying all formulas given above and in Section~\ref{sec:transient_fit} to determine $\eta$, $q$, $\dot M_\text{dust, 1\,au}$, $AD$, $\Delta t$, and $r_\text{nucl}$; and iii) calculating the 16th, 50th, and 84th percentiles of the samples to determine the main values and the 1--$\sigma$ uncertainties given in Table~\ref{tab:kiefer2014}.

\begin{figure}
    \centering
    \includegraphics[width=89.3mm]{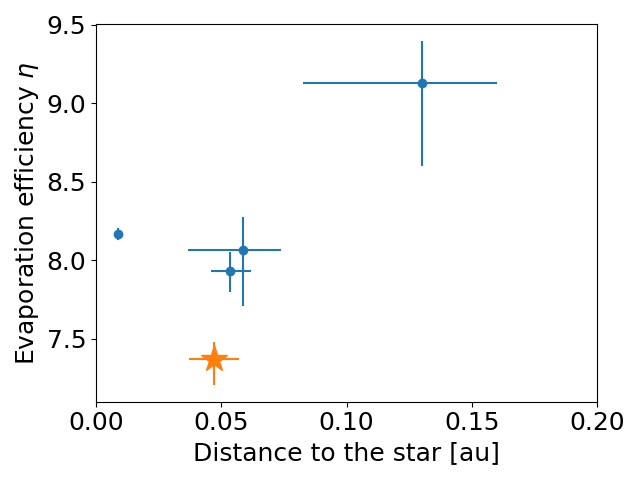}
    \caption{Periastron distance and evaporation efficiency for the exocomets detected in K14 (blue circles) and the candidate exocomet reported in this work (orange star).}
    \label{fig:peri_effi}
\end{figure}

The evaporation efficiency of the CHEOPS exocomet transit reported here can be calculated similarly to eq.~\ref{eq:eff} by using the measured $\dot M_\text{dust,1\,au}$ and fixing $d$=1\,au. We find $\eta$=7.39$\pm$0.14 and an estimated periastron distance of 0.05$\pm$0.01\,au, or equivalently 6.8$\pm$1.4\,$R_\star$. It is graphically compared to the K14 detections in Fig.~\ref{fig:peri_effi}. The CHEOPS transit event and the 2004, 2005, and 2011 K14 events have comparable, though slightly larger, evaporation efficiencies with $\eta_\text{K14}$$\approx$8.05$\pm$$0.10$, but similar periastron distance $q_\text{K14}$$\approx$0.04$\pm$$0.02$\,au. The small difference in evaporation efficiency translates to about four times larger evaporation rates $\dot M_\text{dust, 1\,au}$$\approx$$(4.4$$\pm$$1.0)$$\times$$10^5$\,kg\,s$^{-1}$ and twice as large nucleus radii $\approx$5.0$\pm$0.6\,km. 
In comparison, the 2010 K14 event has radically different properties with $\eta$$>$$8.5$, $q$$>$$0.08$\,au, $\dot M_\text{dust, 1\,au}$$>$$16$$\times$$10^5$\,kg\,s$^{-1}$, and $r_\text{nucl}$$>$$9$\,km. The 2004, 2005, and 2011 K14 spectroscopic detections and the candidate presented in this work hence most likely belong to a common class of exocomets in this system to which the 2010 K14 event would conversely not belong.

The calculated AD and $\Delta t$ for all K14 detections show that if they were observed in photometry, they would have an even larger amplitude in the light curve. 
This implies that the HD\,172555 exocomets that are observed transiting in spectroscopy would produce a detectable photometric signature as well. Given the higher evaporation rates of K14 detections if they are due to a selection effect, the reverse, however, might not always be true. It, nevertheless, opens the opportunity to observe the HD\,172555 exocomets simultaneously with a photometer and a spectrograph in order to detect and measure both the gas and dust evaporated from a single exocomet at once. This could put strong constraints on the relative abundances of chemical elements in exocomets nucleus, that is to say\ volatiles in ices and refractories in dust grains.

\section{Conclusion}
\label{sec:conclusion}
With CHEOPS photometric monitoring of HD\,172555, we have shown that this young star is a $\delta$\,Scuti pulsator. Once the $\delta$\,Scuti oscillations were removed, we discovered a candidate transient event that nicely matched an exocomet transit signature. Fitting the transient light curve with an exocomet transit model, we measured a dust production rate at 1\,au of $\dot M_\text{dust,1au}$$\sim$$10^5$\,kg\,s$^{-1}$. Using the scaling relationship between the dust production rate and nucleus radius~\citep{lecavelier2022}, we deduced that the candidate HD\,172555 exocomet hinted at in this work would have a nucleus size of 2.5$\pm$0.5\,km. This radius is similar to those measured for exocomets in the $\beta$\,Pic system with radii from 1.5 to 6.7\,km~\citep{lecavelier2022}, and for JFC in the Solar System with typical radii of 2--5.5\,km~(\citealt{Tancredi2006}).

More observations on longer uninterrupted time periods of HD\,172555, typically longer than 3\,days, are needed to better characterise the $\delta$\,Scuti variations and more exquisitely remove them from the measured light curve. This is a crucial step to uncover transient absorptions and exclude constructive interferences from residual, low-amplitude $\delta$\,Scuti oscillations. 
Moreover, if the reported transient is indeed a transiting exocomet, we would anticipate with more than 89\% probability a new detection if HD\,172555 is observed on a time interval longer than 4\,days, and a probability of detecting at least two transiting exocomets of at least 73\%. 
After $\beta$\,Pictoris, the young system of HD\,172555 is therefore one of the best targets for investigating the statistics and composition of exocomets, using both spectroscopy and photometry.

\begin{table*}
    \caption{Summary of some parameters from K14's detections. For comparison, the evaporation efficiency obtained for the transient reported here, if due to an exocomet, is $\eta$=7.39$\pm$0.14 for $q$=0.05$\pm$0.01\,au and $\dot M_\text{dust, 1\,au}$=0.92$\pm$0.41$\times$10$^5$\,kg\,s$^{-1}$.}
    \label{tab:kiefer2014}
    \centering
    \begin{tabular}{@{}c@{~~}c@{~~}c@{~~}c@{~~}c@{~~}c@{~~}c@{~~}c@{~~}c@{~~}c@{}}
         Date & $\Delta t_\text{min}$ & $\alpha$ & $A$ & $\eta$ & $q$ & $\dot M_\text{dust, 1\,au}$ & AD & $\Delta t$ & $r_\text{nucl}$\\
              &   &  & & & (au) & ($10^5$\,kg\,s$^{-1}$)&  (ppm) & (hour) & (km)  \\
         \hline
         22/09/2004 & 3.5\,h & $\geq$0.84 & 0.07$\pm$0.005 & 8.17$\pm$0.03 & 0.0088$\pm$0.0006 & 5.6$\pm$0.5 & 5050$\pm$250 & 0.81$\pm$0.03 & 5.66$\pm$0.23  \\
         21/08/2005 & 3\,min & 0.04$^{+0.04}_{-0.01}$ & 1.7$\pm$1.1 & 8.05$\pm$0.30 & 0.059$^{+0.014}_{-0.023}$ &  4.3$\pm$2.5 & 1530$\pm$680 & 2.12$^{+0.24}_{-0.46}$ & 5.0$\pm$1.5  \\
         08/07/2010 & 3\,min & $\geq$0.024 & $\leq$10 & 9.16$^{+0.22}_{-0.57}$ & 0.132$^{+0.026}_{-0.048}$ & 54$\pm$38 & 12800$\pm$7500 & 3.16$^{+0.30}_{-0.64}$ & 17.6$^{+5.1}_{-8.5}$  \\
         11/06/2011 & 1.9\,h & 0.04$\pm$0.01 & 1.7$\pm$0.5 & 7.93$\pm$0.14 & 0.054$^{+0.009}_{-0.010}$ & 3.23$\pm$0.95 & 1210$\pm$280 & 2.02$\pm$0.18 & 4.30$\pm$0.65 \\
         \hline
    \end{tabular}
\end{table*}

\begin{acknowledgements}
CHEOPS is an European Space Agency (ESA) mission in partnership with Switzerland with important contributions to the payload and the ground segment from Austria, Belgium, France, Germany, Hungary, Italy, Portugal, Spain, Sweden, and the United Kingdom. The CHEOPS Consortium would like to gratefully acknowledge the support received by all the agencies, offices, universities, and industries involved. Their flexibility and willingness to explore new approaches were essential to the success of this mission. 
F.K., A.L. \& G.H. acknowledge funding from the Centre National d’Etudes Spatiales and from the French National Research Agency (ANR) under contract number ANR-18-CE31-0019 (SPlaSH). F.K. also acknowledges support from the Université Paris Sciences et Lettres under the DIM-ACAV program Origines et conditions d'apparition de la vie. 
V.V.G.\ is an F.R.S-FNRS Research Associate. 
GyMSz acknowledges the support of the Hungarian National Research, Development and Innovation Office (NKFIH) grant K-125015, a a PRODEX Experiment Agreement No. 4000137122, the Lend\"ulet LP2018-7/2021 grant of the Hungarian Academy of Science and the support of the city of Szombathely. 
ABr was supported by the SNSA. 
ACC acknowledges support from STFC consolidated grant numbers ST/R000824/1 and ST/V000861/1, and UKSA grant number ST/R003203/1. 
ACC and TW acknowledge support from STFC consolidated grant numbers ST/R000824/1 and ST/V000861/1, and UKSA grant number ST/R003203/1. 
S.G.S. acknowledge support from FCT through FCT contract nr. CEECIND/00826/2018 and POPH/FSE (EC).
DG gratefully acknowledges financial support from the CRT foundation under Grant No. 2018.2323 ``Gaseousor rocky? Unveiling the nature of small worlds''. 
YA and MJH acknowledge the support of the Swiss National Fund under grant 200020\_172746. 
We acknowledge support from the Spanish Ministry of Science and Innovation and the European Regional Development Fund through grants ESP2016-80435-C2-1-R, ESP2016-80435-C2-2-R, PGC2018-098153-B-C33, PGC2018-098153-B-C31, ESP2017-87676-C5-1-R, MDM-2017-0737 Unidad de Excelencia Maria de Maeztu-Centro de Astrobiologí­a (INTA-CSIC), as well as the support of the Generalitat de Catalunya/CERCA programme. The MOC activities have been supported by the ESA contract No. 4000124370. 
S.C.C.B.\ acknowledges support from FCT through FCT contracts nr. IF/01312/2014/CP1215/CT0004. 
XB, SC, DG, MF and JL acknowledge their role as ESA-appointed CHEOPS science team members. 
This project was supported by the CNES. 
The Belgian participation to CHEOPS has been supported by the Belgian Federal Science Policy Office (BELSPO) in the framework of the PRODEX Program, and by the University of Liège through an ARC grant for Concerted Research Actions financed by the Wallonia-Brussels Federation. 
L.D. is an F.R.S.-FNRS Postdoctoral Researcher. 
This work was supported by FCT - Fundação para a Ciência e a Tecnologia through national funds and by FEDER through COMPETE2020 - Programa Operacional Competitividade e Internacionalizacão by these grants: UID/FIS/04434/2019, UIDB/04434/2020, UIDP/04434/2020, PTDC/FIS-AST/32113/2017 \& POCI-01-0145-FEDER- 032113, PTDC/FIS-AST/28953/2017 \& POCI-01-0145-FEDER-028953, PTDC/FIS-AST/28987/2017 \& POCI-01-0145-FEDER-028987, O.D.S.D. is supported in the form of work contract (DL 57/2016/CP1364/CT0004) funded by national funds through FCT. 
B.-O.D.\ acknowledges support from the Swiss National Science Foundation (PP00P2-190080). 
This project has received funding from the European Research Council (ERC) under the European Union’s Horizon 2020 research and innovation programme (project {\sc Four Aces}. 
grant agreement No 724427). It has also been carried out in the frame of the National Centre for Competence in Research PlanetS supported by the Swiss National Science Foundation (SNSF). DE acknowledges financial support from the Swiss National Science Foundation for project 200021\_200726. 
MF and CMP gratefully acknowledge the support of the Swedish National Space Agency (DNR 65/19, 174/18). 
M.G.\ is an F.R.S.-FNRS Senior Research Associate. 
SH gratefully acknowledges CNES funding through the grant 837319. 
KGI is the ESA CHEOPS Project Scientist and is responsible for the ESA CHEOPS Guest Observers Programme. She does not participate in, or contribute to, the definition of the Guaranteed Time Programme of the CHEOPS mission through which observations described in this paper have been taken, nor to any aspect of target selection for the programme. 
This work was granted access to the HPC resources of MesoPSL financed by the Region Ile de France and the project Equip@Meso (reference ANR-10-EQPX-29-01) of the programme Investissements d'Avenir supervised by the Agence Nationale pour la Recherche. 
ML acknowledges support of the Swiss National Science Foundation under grant number PCEFP2\_194576. 
PM acknowledges support from STFC research grant number ST/M001040/1. 
LBo, GBr, VNa, IPa, GPi, RRa, GSc, VSi, and TZi acknowledge support from CHEOPS ASI-INAF agreement n. 2019-29-HH.0. 
This work was also partially supported by a grant from the Simons Foundation (PI Queloz, grant number 327127). 
IRI acknowledges support from the Spanish Ministry of Science and Innovation and the European Regional Development Fund through grant PGC2018-098153-B- C33, as well as the support of the Generalitat de Catalunya/CERCA programme. 
NAW acknowledges UKSA grant ST/R004838/1. 
\end{acknowledgements}

\clearpage
\bibliographystyle{aa}
\bibliography{biblio}

\begin{thebibliography}{57}
\expandafter\ifx\csname natexlab\endcsname\relax\def\natexlab#1{#1}\fi

\bibitem[{{Alonso-Floriano} {et~al.}(2015){Alonso-Floriano}, {Caballero},
  {Cort{\'e}s-Contreras}, {Solano}, \& {Montes}}]{Alonso2015}
{Alonso-Floriano}, F.~J., {Caballero}, J.~A., {Cort{\'e}s-Contreras}, M.,
  {Solano}, E., \& {Montes}, D. 2015, \aap, 583, A85

\bibitem[{{Barrado y Navascu{\'e}s} {et~al.}(1999){Barrado y Navascu{\'e}s},
  {Stauffer}, {Song}, \& {Caillault}}]{Barrado1999}
{Barrado y Navascu{\'e}s}, D., {Stauffer}, J.~R., {Song}, I., \& {Caillault},
  J.~P. 1999, \apjl, 520, L123

\bibitem[{{Benz}(2021)}]{Benz2021}
{Benz}, W. 2021, in 43rd COSPAR Scientific Assembly. Held 28 January - 4
  February, Vol.~43, 502

\bibitem[{{Beust} {et~al.}(1990){Beust}, {Lagrange-Henri}, {Vidal-Madjar}, \&
  {Ferlet}}]{Beust1990}
{Beust}, H., {Lagrange-Henri}, A.~M., {Vidal-Madjar}, A., \& {Ferlet}, R. 1990,
  \aap, 236, 202

\bibitem[{{Beust} \& {Tagger}(1993)}]{Beust1993}
{Beust}, H. \& {Tagger}, M. 1993, \icarus, 106, 42

\bibitem[{{Binks} \& {Jeffries}(2014)}]{Binks2014}
{Binks}, A.~S. \& {Jeffries}, R.~D. 2014, \mnras, 438, L11

\bibitem[{{Blackwell} \& {Shallis}(1977)}]{Blackwell1977}
{Blackwell}, D.~E. \& {Shallis}, M.~J. 1977, \mnras, 180, 177

\bibitem[{{Bonfanti} {et~al.}(2021){Bonfanti}, {Delrez}, {Hooton}, {Wilson},
  {Fossati}, {Alibert}, {Hoyer}, {Mustill}, {Osborn}, {Adibekyan}, {Gandolfi},
  {Salmon}, {Sousa}, {Tuson}, {Van Grootel}, {Cabrera}, {Nascimbeni}, {Maxted},
  {Barros}, {Billot}, {Bonfils}, {Borsato}, {Broeg}, {Davies}, {Deleuil},
  {Demangeon}, {Fridlund}, {Lacedelli}, {Lendl}, {Persson}, {Santos},
  {Scandariato}, {Szab{\'o}}, {Collier Cameron}, {Udry}, {Benz}, {Beck},
  {Ehrenreich}, {Fortier}, {Isaak}, {Queloz}, {Alonso}, {Asquier}, {Bandy},
  {B{\'a}rczy}, {Barrado}, {Barrag{\'a}n}, {Baumjohann}, {Beck}, {Bekkelien},
  {Bergomi}, {Brandeker}, {Busch}, {Cessa}, {Charnoz}, {Chazelas}, {Corral Van
  Damme}, {Demory}, {Erikson}, {Farinato}, {Futyan}, {Garcia Mu{\~n}oz},
  {Gillon}, {Guedel}, {Guterman}, {Hasiba}, {Heng}, {Hernandez}, {Kiss},
  {Kuntzer}, {Laskar}, {Lecavelier des Etangs}, {Lovis}, {Magrin}, {Malvasio},
  {Marafatto}, {Michaelis}, {Munari}, {Olofsson}, {Ottacher}, {Ottensamer},
  {Pagano}, {Pall{\'e}}, {Peter}, {Piazza}, {Piotto}, {Pollacco}, {Ragazzoni},
  {Rando}, {Ratti}, {Rauer}, {Ribas}, {Rieder}, {Rohlfs}, {Safa}, {Salatti},
  {S{\'e}gransan}, {Simon}, {Smith}, {Sordet}, {Steller}, {Thomas},
  {Tschentscher}, {Van Eylen}, {Viotto}, {Walter}, {Walton}, {Wildi}, \&
  {Wolter}}]{Bonfanti2021}
{Bonfanti}, A., {Delrez}, L., {Hooton}, M.~J., {et~al.} 2021, \aap, 646, A157

\bibitem[{{Bottke} {et~al.}(2012){Bottke}, {Vokrouhlick{\'y}}, {Minton},
  {Nesvorn{\'y}}, {Morbidelli}, {Brasser}, {Simonson}, \&
  {Levison}}]{Bottke2012}
{Bottke}, W.~F., {Vokrouhlick{\'y}}, D., {Minton}, D., {et~al.} 2012, \nat,
  485, 78

\bibitem[{{Brandeker} {et~al.}(2022){Brandeker}, {Heng}, {Lendl}, {Patel},
  {Morris}, {Broeg}, {Guterman}, {Beck}, {Maxted}, {Demangeon}, {Delrez},
  {Demory}, {Kitzmann}, {Santos}, {Singh}, {Alibert}, {Alonso}, {Anglada},
  {B{\'a}rczy}, {Barrado y Navascues}, {Barros}, {Baumjohann}, {Beck}, {Benz},
  {Billot}, {Bonfils}, {Bruno}, {Cabrera}, {Charnoz}, {Collier Cameron},
  {Corral van Damme}, {Csizmadia}, {Davies}, {Deleuil}, {Deline}, {Ehrenreich},
  {Erikson}, {Farinato}, {Fortier}, {Fossati}, {Fridlund}, {Gandolfi},
  {Gillon}, {G{\"u}del}, {Hoyer}, {Isaak}, {Kiss}, {Laskar}, {Lecavelier des
  Etangs}, {Lovis}, {Luntzer}, {Magrin}, {Nascimbeni}, {Olofsson},
  {Ottensamer}, {Pagano}, {Pall{\'e}}, {Peter}, {Piotto}, {Pollacco}, {Queloz},
  {Ragazzoni}, {Rando}, {Rauer}, {Ribas}, {Scandariato}, {S{\'e}gransan},
  {Simon}, {Smith}, {Sousa}, {Steller}, {Szab{\'o}}, {Thomas}, {Udry}, {Van
  Grootel}, {Walton}, \& {Wolter}}]{Brandeker2022}
{Brandeker}, A., {Heng}, K., {Lendl}, M., {et~al.} 2022, \aap, 659, L4

\bibitem[{{Cartwright} {et~al.}(2022){Cartwright}, {Hodges}, \&
  {Wadhwa}}]{Cartwright2022}
{Cartwright}, J.~A., {Hodges}, K.~V., \& {Wadhwa}, M. 2022, Earth and Planetary
  Science Letters, 590, 117576

\bibitem[{{Chambers} \& {Lissauer}(2002)}]{chambers2002}
{Chambers}, J.~E. \& {Lissauer}, J.~J. 2002, in Lunar and Planetary Science
  Conference, Lunar and Planetary Science Conference, 1093

\bibitem[{{Charpinet} {et~al.}(2010){Charpinet}, {Green}, {Baglin}, {Van
  Grootel}, {Fontaine}, {Vauclair}, {Chaintreuil}, {Weiss}, {Michel},
  {Auvergne}, {Catala}, {Samadi}, \& {Baudin}}]{Charpinet2010}
{Charpinet}, S., {Green}, E.~M., {Baglin}, A., {et~al.} 2010, \aap, 516, L6

\bibitem[{{de Sousa} {et~al.}(2020){de Sousa}, {Morbidelli}, {Raymond},
  {Izidoro}, {Gomes}, \& {Vieira Neto}}]{desousa2020}
{de Sousa}, R.~R., {Morbidelli}, A., {Raymond}, S.~N., {et~al.} 2020, \icarus,
  339, 113605

\bibitem[{{Deeming}(1975)}]{Deeming-75}
{Deeming}, T.~J. 1975, \apss, 36, 137

\bibitem[{{Deline} {et~al.}(2022){Deline}, {Hooton}, {Lendl}, {Morris},
  {Salmon}, {Olofsson}, {Broeg}, {Ehrenreich}, {Beck}, {Brandeker}, {Hoyer},
  {Sulis}, {Van Grootel}, {Bourrier}, {Demangeon}, {Demory}, {Heng},
  {Parviainen}, {Serrano}, {Singh}, {Bonfanti}, {Fossati}, {Kitzmann}, {Sousa},
  {Wilson}, {Alibert}, {Alonso}, {Anglada}, {B{\'a}rczy}, {Barrado Navascues},
  {Barros}, {Baumjohann}, {Beck}, {Bekkelien}, {Benz}, {Billot}, {Bonfils},
  {Cabrera}, {Charnoz}, {Collier Cameron}, {Corral van Damme}, {Csizmadia},
  {Davies}, {Deleuil}, {Delrez}, {de Roche}, {Erikson}, {Fortier}, {Fridlund},
  {Futyan}, {Gandolfi}, {Gillon}, {G{\"u}del}, {Gutermann}, {Hasiba}, {Isaak},
  {Kiss}, {Laskar}, {Lecavelier des Etangs}, {Lovis}, {Magrin}, {Maxted},
  {Munari}, {Nascimbeni}, {Ottensamer}, {Pagano}, {Pall{\'e}}, {Peter},
  {Piotto}, {Pollacco}, {Queloz}, {Ragazzoni}, {Rando}, {Rauer}, {Ribas},
  {Santos}, {Scandariato}, {S{\'e}gransan}, {Simon}, {Smith}, {Steller},
  {Szab{\'o}}, {Thomas}, {Udry}, {Walter}, \& {Walton}}]{deline2022}
{Deline}, A., {Hooton}, M.~J., {Lendl}, M., {et~al.} 2022, \aap, 659, A74

\bibitem[{{Engler} {et~al.}(2018){Engler}, {Schmid}, {Quanz}, {Avenhaus}, \&
  {Bazzon}}]{Engler2018}
{Engler}, N., {Schmid}, H.~M., {Quanz}, S.~P., {Avenhaus}, H., \& {Bazzon}, A.
  2018, \aap, 618, A151

\bibitem[{{Erspamer} \& {North}(2003)}]{Erspamer2003}
{Erspamer}, D. \& {North}, P. 2003, \aap, 398, 1121

\bibitem[{{Feigelson} {et~al.}(2006){Feigelson}, {Lawson}, {Stark}, {Townsley},
  \& {Garmire}}]{Feigelson2006}
{Feigelson}, E.~D., {Lawson}, W.~A., {Stark}, M., {Townsley}, L., \& {Garmire},
  G.~P. 2006, \aj, 131, 1730

\bibitem[{{Ferlet} {et~al.}(1987){Ferlet}, {Hobbs}, \&
  {Vidal-Madjar}}]{Ferlet1987}
{Ferlet}, R., {Hobbs}, L.~M., \& {Vidal-Madjar}, A. 1987, \aap, 185, 267

\bibitem[{{Gaia Collaboration}(2020)}]{Gaia2020}
{Gaia Collaboration}. 2020, VizieR Online Data Catalog, I/350

\bibitem[{{Gomes} {et~al.}(2005){Gomes}, {Levison}, {Tsiganis}, \&
  {Morbidelli}}]{Gomes2005}
{Gomes}, R., {Levison}, H.~F., {Tsiganis}, K., \& {Morbidelli}, A. 2005, \nat,
  435, 466

\bibitem[{{Grady} {et~al.}(2018){Grady}, {Brown}, {Welsh}, {Roberge}, {Kamp},
  \& {Rivi{\`e}re Marichalar}}]{Grady2018}
{Grady}, C.~A., {Brown}, A., {Welsh}, B., {et~al.} 2018, \aj, 155, 242

\bibitem[{{Gray} {et~al.}(2006){Gray}, {Corbally}, {Garrison}, {McFadden},
  {Bubar}, {McGahee}, {O'Donoghue}, \& {Knox}}]{Gray2006}
{Gray}, R.~O., {Corbally}, C.~J., {Garrison}, R.~F., {et~al.} 2006, \aj, 132,
  161

\bibitem[{{H{\o}g} {et~al.}(2000){H{\o}g}, {Fabricius}, {Makarov}, {Urban},
  {Corbin}, {Wycoff}, {Bastian}, {Schwekendiek}, \& {Wicenec}}]{Hog2000}
{H{\o}g}, E., {Fabricius}, C., {Makarov}, V.~V., {et~al.} 2000, \aap, 355, L27

\bibitem[{{Hoyer} {et~al.}(2020){Hoyer}, {Guterman}, {Demangeon}, {Sousa},
  {Deleuil}, {Meunier}, \& {Benz}}]{hoyer2020}
{Hoyer}, S., {Guterman}, P., {Demangeon}, O., {et~al.} 2020, \aap, 635, A24

\bibitem[{{Jewitt} \& {Matthews}(1999)}]{Jewitt1999}
{Jewitt}, D. \& {Matthews}, H. 1999, \aj, 117, 1056

\bibitem[{{Kiefer} {et~al.}(2014{\natexlab{a}}){Kiefer}, {Lecavelier des
  Etangs}, {Augereau}, {Vidal-Madjar}, {Lagrange}, \& {Beust}}]{kiefer2014a}
{Kiefer}, F., {Lecavelier des Etangs}, A., {Augereau}, J.~C., {et~al.}
  2014{\natexlab{a}}, \aap, 561, L10

\bibitem[{{Kiefer} {et~al.}(2014{\natexlab{b}}){Kiefer}, {Lecavelier des
  Etangs}, {Boissier}, {Vidal-Madjar}, {Beust}, {Lagrange}, {H{\'e}brard}, \&
  {Ferlet}}]{Kiefer2014b}
{Kiefer}, F., {Lecavelier des Etangs}, A., {Boissier}, J., {et~al.}
  2014{\natexlab{b}}, \nat, 514, 462

\bibitem[{{Kiefer} {et~al.}(2017){Kiefer}, {Lecavelier des {\'E}tangs},
  {Vidal-Madjar}, {H{\'e}brard}, {Bourrier}, \& {Wilson}}]{Kiefer2017}
{Kiefer}, F., {Lecavelier des {\'E}tangs}, A., {Vidal-Madjar}, A., {et~al.}
  2017, \aap, 608, A132

\bibitem[{{Lagrange-Henri} {et~al.}(1992){Lagrange-Henri}, {Gosset}, {Beust},
  {Ferlet}, \& {Vidal-Madjar}}]{Lagrange1992}
{Lagrange-Henri}, A.~M., {Gosset}, E., {Beust}, H., {Ferlet}, R., \&
  {Vidal-Madjar}, A. 1992, \aap, 264, 637

\bibitem[{{Lecavelier des Etangs} {et~al.}(2022){Lecavelier des Etangs},
  {Cros}, {H{\'e}brard}, {Martioli}, {Duquesnoy}, {Kenworthy}, {Kiefer},
  {Lacour}, {Lagrange}, {Meunier}, \& {Vidal-Madjar}}]{lecavelier2022}
{Lecavelier des Etangs}, A., {Cros}, L., {H{\'e}brard}, G., {et~al.} 2022,
  Scientific Reports, 12, 5855

\bibitem[{{Lecavelier Des Etangs} {et~al.}(1999){Lecavelier Des Etangs},
  {Vidal-Madjar}, \& {Ferlet}}]{Lecavelier1999}
{Lecavelier Des Etangs}, A., {Vidal-Madjar}, A., \& {Ferlet}, R. 1999, \aap,
  343, 916

\bibitem[{{Lisse} {et~al.}(2009){Lisse}, {Chen}, {Wyatt}, {Morlok}, {Song},
  {Bryden}, \& {Sheehan}}]{Lisse2009}
{Lisse}, C.~M., {Chen}, C.~H., {Wyatt}, M.~C., {et~al.} 2009, \apj, 701, 2019

\bibitem[{{Liu} {et~al.}(2022){Liu}, {Raymond}, \& {Jacobson}}]{beibei2022}
{Liu}, B., {Raymond}, S.~N., \& {Jacobson}, S.~A. 2022, \nat, 604, 643

\bibitem[{{Mamajek} \& {Bell}(2014)}]{Mamajek2014}
{Mamajek}, E.~E. \& {Bell}, C. P.~M. 2014, \mnras, 445, 2169

\bibitem[{{Marks} {et~al.}(2019){Marks}, {Borg}, {Shearer}, \&
  {Cassata}}]{Marks2019}
{Marks}, N.~E., {Borg}, L.~E., {Shearer}, C.~K., \& {Cassata}, W.~S. 2019,
  Journal of Geophysical Research (Planets), 124, 2465

\bibitem[{{Miret-Roig} {et~al.}(2018){Miret-Roig}, {Antoja},
  {Romero-G{\'o}mez}, \& {Figueras}}]{Miret2018}
{Miret-Roig}, N., {Antoja}, T., {Romero-G{\'o}mez}, M., \& {Figueras}, F. 2018,
  \aap, 615, A51

\bibitem[{{Morbidelli} {et~al.}(2018){Morbidelli}, {Nesvorny}, {Laurenz},
  {Marchi}, {Rubie}, {Elkins-Tanton}, {Wieczorek}, \& {Jacobson}}]{morbi2018}
{Morbidelli}, A., {Nesvorny}, D., {Laurenz}, V., {et~al.} 2018, \icarus, 305,
  262

\bibitem[{{Morbidelli} {et~al.}(2001){Morbidelli}, {Petit}, {Gladman}, \&
  {Chambers}}]{morbi2001}
{Morbidelli}, A., {Petit}, J.~M., {Gladman}, B., \& {Chambers}, J. 2001, maps,
  36, 371

\bibitem[{{Nilsson} {et~al.}(2009){Nilsson}, {Liseau}, {Brandeker}, {Olofsson},
  {Risacher}, {Fridlund}, \& {Pilbratt}}]{Nilsson2009}
{Nilsson}, R., {Liseau}, R., {Brandeker}, A., {et~al.} 2009, \aap, 508, 1057

\bibitem[{{Pavlenko} {et~al.}(2022){Pavlenko}, {Kulyk}, {Shubina}, {Vasylenko},
  {Dobrycheva}, \& {Korsun}}]{Pavlenko2022}
{Pavlenko}, Y., {Kulyk}, I., {Shubina}, O., {et~al.} 2022, \aap, 660, A49

\bibitem[{{Rappaport} {et~al.}(2018){Rappaport}, {Vanderburg}, {Jacobs},
  {LaCourse}, {Jenkins}, {Kraus}, {Rizzuto}, {Latham}, {Bieryla}, {Lazarevic},
  \& {Schmitt}}]{Rappaport2018}
{Rappaport}, S., {Vanderburg}, A., {Jacobs}, T., {et~al.} 2018, \mnras, 474,
  1453

\bibitem[{{Riviere-Marichalar} {et~al.}(2012){Riviere-Marichalar}, {Barrado},
  {Augereau}, {Thi}, {Roberge}, {Eiroa}, {Montesinos}, {Meeus}, {Howard},
  {Sandell}, {Duch{\^e}ne}, {Dent}, {Lebreton}, {Mendigut{\'\i}a},
  {Hu{\'e}lamo}, {M{\'e}nard}, \& {Pinte}}]{riviere2012}
{Riviere-Marichalar}, P., {Barrado}, D., {Augereau}, J.~C., {et~al.} 2012,
  \aap, 546, L8

\bibitem[{{Ryder}(1990)}]{ryder1990}
{Ryder}, G. 1990, EOS Transactions, 71, 313

\bibitem[{{Schanche} {et~al.}(2020){Schanche}, {H{\'e}brard}, {Collier
  Cameron}, {Dalal}, {Smalley}, {Wilson}, {Boisse}, {Bouchy}, {Brown},
  {Demangeon}, {Haswell}, {Hellier}, {Kolb}, {Lopez}, {Maxted}, {Pollacco},
  {West}, \& {Wheatley}}]{Schanche2020}
{Schanche}, N., {H{\'e}brard}, G., {Collier Cameron}, A., {et~al.} 2020,
  \mnras, 499, 428

\bibitem[{{Schneiderman} {et~al.}(2021){Schneiderman}, {Matr{\`a}}, {Jackson},
  {Kennedy}, {Kral}, {Marino}, {{\"O}berg}, {Su}, {Wilner}, \&
  {Wyatt}}]{Schneiderman2021}
{Schneiderman}, T., {Matr{\`a}}, L., {Jackson}, A.~P., {et~al.} 2021, \nat,
  598, 425

\bibitem[{{Smith} {et~al.}(2012){Smith}, {Wyatt}, \& {Haniff}}]{Smith2012}
{Smith}, R., {Wyatt}, M.~C., \& {Haniff}, C.~A. 2012, \mnras, 422, 2560

\bibitem[{{Szab{\'o}} {et~al.}(2021){Szab{\'o}}, {Gandolfi}, {Brandeker},
  {Csizmadia}, {Garai}, {Billot}, {Broeg}, {Ehrenreich}, {Fortier}, {Fossati},
  {Hoyer}, {Kiss}, {Lecavelier des Etangs}, {Maxted}, {Ribas}, {Alibert},
  {Alonso}, {Anglada Escud{\'e}}, {B{\'a}rczy}, {Barros}, {Barrado},
  {Baumjohann}, {Beck}, {Beck}, {Bekkelien}, {Bonfils}, {Benz}, {Borsato},
  {Busch}, {Cabrera}, {Charnoz}, {Collier Cameron}, {Van Damme}, {Davies},
  {Delrez}, {Deleuil}, {Demangeon}, {Demory}, {Erikson}, {Fridlund}, {Futyan},
  {Garc{\'\i}a Mu{\~n}oz}, {Gillon}, {Guedel}, {Guterman}, {Heng}, {Isaak},
  {Lacedelli}, {Laskar}, {Lendl}, {Lovis}, {Luntzer}, {Magrin}, {Nascimbeni},
  {Olofsson}, {Osborn}, {Ottensamer}, {Pagano}, {Pall{\'e}}, {Peter}, {Piazza},
  {Piotto}, {Pollacco}, {Queloz}, {Ragazzoni}, {Rando}, {Rauer}, {Santos},
  {Scandariato}, {S{\'e}gransan}, {Serrano}, {Sicilia}, {Simon}, {Smith},
  {Sousa}, {Steller}, {Thomas}, {Udry}, {Van Grootel}, {Walton}, \&
  {Wilson}}]{szabo2021}
{Szab{\'o}}, G.~M., {Gandolfi}, D., {Brandeker}, A., {et~al.} 2021, \aap, 654,
  A159

\bibitem[{{Tancredi} {et~al.}(2006){Tancredi}, {Fern{\'a}ndez}, {Rickman}, \&
  {Licandro}}]{Tancredi2006}
{Tancredi}, G., {Fern{\'a}ndez}, J.~A., {Rickman}, H., \& {Licandro}, J. 2006,
  \icarus, 182, 527

\bibitem[{{Tsiganis} {et~al.}(2005){Tsiganis}, {Gomes}, {Morbidelli}, \&
  {Levison}}]{Tsiganis2005}
{Tsiganis}, K., {Gomes}, R., {Morbidelli}, A., \& {Levison}, H.~F. 2005, \nat,
  435, 459

\bibitem[{{Uytterhoeven} {et~al.}(2011){Uytterhoeven}, {Moya},
  {Grigahc{\`e}ne}, {Guzik}, {Guti{\'e}rrez-Soto}, {Smalley}, {Handler},
  {Balona}, {Niemczura}, {Fox Machado}, {Benatti}, {Chapellier}, {Tkachenko},
  {Szab{\'o}}, {Su{\'a}rez}, {Ripepi}, {Pascual}, {Mathias},
  {Mart{\'\i}n-Ru{\'\i}z}, {Lehmann}, {Jackiewicz}, {Hekker}, {Gruberbauer},
  {Garc{\'\i}a}, {Dumusque}, {D{\'\i}az-Fraile}, {Bradley}, {Antoci}, {Roth},
  {Leroy}, {Murphy}, {De Cat}, {Cuypers}, {Kjeldsen}, {Christensen-Dalsgaard},
  {Breger}, {Pigulski}, {Kiss}, {Still}, {Thompson}, \& {van
  Cleve}}]{Uytterhoeven2011}
{Uytterhoeven}, K., {Moya}, A., {Grigahc{\`e}ne}, A., {et~al.} 2011, \aap, 534,
  A125

\bibitem[{{Vidal-Madjar} {et~al.}(1994){Vidal-Madjar}, {Lagrange-Henri},
  {Feldman}, {Beust}, {Lissauer}, {Deleuil}, {Ferlet}, {Gry}, {Hobbs},
  {McGrath}, {McPhate}, \& {Moos}}]{Vidal1994}
{Vidal-Madjar}, A., {Lagrange-Henri}, A.~M., {Feldman}, P.~D., {et~al.} 1994,
  \aap, 290, 245

\bibitem[{{Wilson} {et~al.}(2022){Wilson}, {Goffo}, {Alibert}, {Gandolfi},
  {Bonfanti}, {Persson}, {Collier Cameron}, {Fridlund}, {Fossati}, {Korth},
  {Benz}, {Deline}, {Flor{\'e}n}, {Guterman}, {Adibekyan}, {Hooton}, {Hoyer},
  {Leleu}, {Mustill}, {Salmon}, {Sousa}, {Suarez}, {Abe}, {Agabi}, {Alonso},
  {Anglada}, {Asquier}, {B{\'a}rczy}, {Barrado Navascues}, {Barros},
  {Baumjohann}, {Beck}, {Beck}, {Billot}, {Bonfils}, {Brandeker}, {Broeg},
  {Bryant}, {Burleigh}, {Buttu}, {Cabrera}, {Charnoz}, {Ciardi}, {Cloutier},
  {Cochran}, {Collins}, {Col{\'o}n}, {Crouzet}, {Csizmadia}, {Davies},
  {Deleuil}, {Delrez}, {Demangeon}, {Demory}, {Dragomir}, {Dransfield},
  {Ehrenreich}, {Erikson}, {Fortier}, {Gan}, {Gill}, {Gillon}, {Gnilka},
  {Grieves}, {Grziwa}, {G{\"u}del}, {Guillot}, {Haldemann}, {Heng}, {Horne},
  {Howell}, {Isaak}, {Jenkins}, {Jensen}, {Kiss}, {Lacedelli}, {Lam}, {Laskar},
  {Latham}, {Lecavelier des Etangs}, {Lendl}, {Lester}, {Levine}, {Livingston},
  {Lovis}, {Luque}, {Magrin}, {Marie-Sainte}, {Maxted}, {Mayo}, {McLean},
  {Mecina}, {M{\'e}karnia}, {Nascimbeni}, {Nielsen}, {Olofsson}, {Osborn},
  {Osborne}, {Ottensamer}, {Pagano}, {Pall{\'e}}, {Peter}, {Piotto},
  {Pollacco}, {Queloz}, {Ragazzoni}, {Rando}, {Rauer}, {Redfield}, {Ribas},
  {Ricker}, {Rieder}, {Santos}, {Scandariato}, {Schmider}, {Schwarz}, {Scott},
  {Seager}, {S{\'e}gransan}, {Serrano}, {Simon}, {Smith}, {Steller},
  {Stockdale}, {Szab{\'o}}, {Thomas}, {Ting}, {Triaud}, {Udry}, {Van Eylen},
  {Van Grootel}, {Vanderspek}, {Viotto}, {Walton}, \& {Winn}}]{Wilson2022}
{Wilson}, T.~G., {Goffo}, E., {Alibert}, Y., {et~al.} 2022, \mnras, 511, 1043

\bibitem[{{Zieba} {et~al.}(2019){Zieba}, {Zwintz}, {Kenworthy}, \&
  {Kennedy}}]{Zieba2019}
{Zieba}, S., {Zwintz}, K., {Kenworthy}, M.~A., \& {Kennedy}, G.~M. 2019, \aap,
  625, L13

\bibitem[{{Zong} {et~al.}(2016){Zong}, {Charpinet}, \& {Vauclair}}]{Zong2016}
{Zong}, W., {Charpinet}, S., \& {Vauclair}, G. 2016, \aap, 594, A46

\bibitem[{{Zuckerman} {et~al.}(2001){Zuckerman}, {Song}, {Bessell}, \&
  {Webb}}]{Zuckerman2001}
{Zuckerman}, B., {Song}, I., {Bessell}, M.~S., \& {Webb}, R.~A. 2001, \apjl,
  562, L87

\end{thebibliography}

\clearpage
\onecolumn
            
\appendix

\section{List of periodic variabilities in the CHEOPS light curve of HD\,172555}

\footnotesize
\begin{longtable}{lccccccccccl}
\caption{List of frequencies $f_{n}$ extracted in the CHEOPS light curve of HD\,172555 down to S/N=4.8 and their fitted parameters. The CHEOPS orbital frequency and its harmonics are indicated in the 'Comments' column.\label{app:Freq}}\\
\hline\hline
Id. &   & Frequency  & $\sigma_f$ & Period & $\sigma_P$    & Amplitude & $\sigma_A$ & Phase & $\sigma_{\rm Ph}$ & S/N & Comments\tabularnewline
    &   & ($\mu$Hz) & ($\mu$Hz)  & (s)    & (s)            & ($\%$)  & ($\%$)    &       &                     &     &         \tabularnewline
\hline

 & \tabularnewline
\endfirsthead
\caption{continued.}\\
\hline\hline
Id. &   & Frequency  & $\sigma_f$ & Period & $\sigma_P$    & Amplitude & $\sigma_A$ & Phase & $\sigma_{\rm Ph}$ & S/N & Comments\tabularnewline
    &   & ($\mu$Hz) & ($\mu$Hz)  & (s)    & (s)            & ($\%$)  & ($\%$)    &       &                     &     &         \tabularnewline
\hline

 & \tabularnewline
\endhead

 & \tabularnewline
\hline

\endfoot

 & \tabularnewline
\hline

\endlastfoot

$f_{39}$  & & $103.41$   & $0.58$   & $9670.24$   & $55$   & $0.0060$   & $0.0011$   & $0.740$   & $0.060$   & $5.3$  &  \tabularnewline
$f_{14}$  & & $169.48$   & $0.17$   & $5900.40$   & $6.0$   & $0.0198$   & $0.0011$   & $0.215$   & $0.018$   & $18.0$   & $\sim f_{\rm orb}$ \tabularnewline
$f_{29}$  & & $193.29$   & $0.36$   & $5173.57$   & $9.7$   & $0.0093$   & $0.0011$   & $0.469$   & $0.037$   & $8.5$   &  \tabularnewline
$f_{7}$  & & $310.588$   & $0.070$   & $3219.70$   & $0.73$   & $0.0460$   & $0.0010$   & $0.8054$   & $0.0072$   & $44.1$   & \tabularnewline
$f_{21}$  & & $358.91$   & $0.27$   & $2786.21$   & $2.1$   & $0.0118$   & $0.0010$   & $0.960$   & $0.028$   & $11.4$   &  \tabularnewline
$f_{4}$  & & $387.405$   & $0.050$   & $2581.28$   & $0.33$   & $0.0641$   & $0.0010$   & $0.7624$   & $0.0051$   & $62.5$   &  \tabularnewline
$f_{10}$  & & $391.669$   & $0.094$   & $2553.17$   & $0.61$   & $0.0339$   & $0.0010$   & $0.6527$   & $0.0096$   & $33.2$   &  \tabularnewline
$f_{24}$  & & $396.21$   & $0.31$   & $2523.90$   & $2.0$   & $0.0103$   & $0.0010$   & $0.862$   & $0.032$   & $10.1$   &  \tabularnewline
$f_{8}$  & & $409.295$   & $0.092$   & $2443.22$   & $0.55$   & $0.0341$   & $0.0010$   & $0.5535$   & $0.0094$   & $33.7$   &  \tabularnewline
$f_{1}$  & & $423.084$   & $0.023$   & $2363.60$   & $0.13$   & $0.1340$   & $0.0010$   & $0.7407$   & $0.0024$   & $133.0$   &  \tabularnewline
$f_{13}$  & & $439.82$   & $0.15$   & $2273.64$   & $0.76$   & $0.0212$   & $0.0010$   & $0.504$   & $0.015$   & $21.1$   &  \tabularnewline
$f_{20}$  & & $452.69$   & $0.23$   & $2209.00$   & $1.1$   & $0.01344$   & $0.00100$   & $0.869$   & $0.024$   & $13.5$   &  \tabularnewline
$f_{3}$  & & $458.138$   & $0.044$   & $2182.75$   & $0.21$   & $0.07081$   & $0.00100$   & $0.1308$   & $0.0045$   & $71.1$   &  \tabularnewline
$f_{11}$  & & $465.88$   & $0.10$   & $2146.48$   & $0.47$   & $0.03035$   & $0.00099$   & $0.139$   & $0.010$   & $30.6$   &  \tabularnewline
$f_{25}$  & & $474.33$   & $0.30$   & $2108.2$   & $1.3$   & $0.01021$   & $0.00099$   & $0.649$   & $0.031$   & $10.3$   &  \tabularnewline
$f_{6}$  & & $480.561$   & $0.060$   & $2080.90$   & $0.26$   & $0.05146$   & $0.00099$   & $0.4813$   & $0.0061$   & $52.1$   &  \tabularnewline
$f_{12}$  & & $486.26$   & $0.10$   & $2056.52$   & $0.43$   & $0.03021$   & $0.00099$   & $0.763$   & $0.010$   & $30.6$   &  \tabularnewline
$f_{22}$  & & $490.93$   & $0.26$   & $2036.90$   & $1.1$   & $0.01159$   & $0.00098$   & $0.089$   & $0.027$   & $11.8$   &  \tabularnewline
$f_{5}$  & & $501.348$   & $0.058$   & $1994.62$   & $0.23$   & $0.05218$   & $0.00098$   & $0.2455$   & $0.0060$   & $53.3$   &  \tabularnewline
$f_{15}$  & & $520.98$   & $0.16$   & $1919.44$   & $0.59$   & $0.01888$   & $0.00097$   & $0.450$   & $0.016$   & $19.4$   &  \tabularnewline
$f_{2}$  & & $536.714$   & $0.035$   & $1863.19$   & $0.12$   & $0.08470$   & $0.00096$   & $0.2276$   & $0.0036$   & $87.9$   &  \tabularnewline
$f_{9}$  & & $548.173$   & $0.088$   & $1824.24$   & $0.29$   & $0.03397$   & $0.00096$   & $0.4118$   & $0.0090$   & $35.3$   &  \tabularnewline
$f_{27}$  & & $574.38$   & $0.31$   & $1741.01$   & $0.95$   & $0.00947$   & $0.00096$   & $0.811$   & $0.032$   & $9.9$   &  \tabularnewline
$f_{19}$  & & $583.23$   & $0.20$   & $1714.58$   & $0.58$   & $0.01501$   & $0.00095$   & $0.354$   & $0.020$   & $15.8$   &  \tabularnewline
$f_{30}$  & & $597.04$   & $0.32$   & $1674.92$   & $0.90$   & $0.00916$   & $0.00095$   & $0.130$   & $0.033$   & $9.7$   &  \tabularnewline
$f_{26}$  & & $609.30$   & $0.30$   & $1641.22$   & $0.81$   & $0.00969$   & $0.00094$   & $0.039$   & $0.031$   & $10.3$   & \tabularnewline
$f_{31}$  & & $663.47$   & $0.33$   & $1507.24$   & $0.75$   & $0.00870$   & $0.00093$   & $0.964$   & $0.034$   & $9.4$   &  \tabularnewline
$f_{18}$  & & $673.06$   & $0.19$   & $1485.76$   & $0.42$   & $0.01508$   & $0.00092$   & $0.969$   & $0.019$   & $16.4$   & $\sim 4*f_{\rm orb}$ \tabularnewline
$f_{40}$  & & $676.16$   & $0.49$   & $1478.90$   & $1.1$   & $0.00580$   & $0.00092$   & $0.056$   & $0.050$   & $6.3$   &  \tabularnewline
$f_{33}$  & & $691.33$   & $0.33$   & $1446.49$   & $0.70$   & $0.00852$   & $0.00092$   & $0.904$   & $0.034$   & $9.3$   &  \tabularnewline
$f_{32}$  & & $700.44$   & $0.33$   & $1427.67$   & $0.68$   & $0.00852$   & $0.00091$   & $0.054$   & $0.034$   & $9.4$   &  \tabularnewline
$f_{35}$  & & $783.29$   & $0.36$   & $1276.67$   & $0.58$   & $0.00770$   & $0.00089$   & $0.718$   & $0.037$   & $8.7$   &  \tabularnewline
$f_{43}$  & & $815.85$   & $0.52$   & $1225.72$   & $0.79$   & $0.00519$   & $0.00088$   & $0.739$   & $0.054$   & $5.9$   & \tabularnewline
$f_{42}$  & & $894.11$   & $0.50$   & $1118.44$   & $0.63$   & $0.00531$   & $0.00086$   & $0.326$   & $0.052$   & $6.2$   & \tabularnewline
$f_{34}$  & & $1011.23$   & $0.31$   & $988.89$   & $0.30$   & $0.00832$   & $0.00083$   & $0.782$   & $0.032$   & $10.0$   & 6*$f_{\rm orb}$ \tabularnewline
$f_{44}$  & & $1180.09$   & $0.61$   & $847.39$   & $0.44$   & $0.00399$   & $0.00079$   & $0.174$   & $0.063$   & $5.1$   & 7*$f_{\rm orb}$ \tabularnewline
$f_{41}$  & & $1685.15$   & $0.41$   & $593.42$   & $0.14$   & $0.00540$   & $0.00071$   & $0.746$   & $0.042$   & $7.6$   & 10*$f_{\rm orb}$ \tabularnewline
$f_{38}$  & & $2190.70$   & $0.30$   & $456.48$   & $0.062$   & $0.00628$   & $0.00060$   & $0.122$   & $0.030$   & $10.5$   & 13*$f_{\rm orb}$ \tabularnewline
$f_{36}$  & & $2527.68$   & $0.25$   & $395.62$   & $0.039$   & $0.00670$   & $0.00054$   & $0.293$   & $0.026$   & $12.3$   & 15*$f_{\rm orb}$ \tabularnewline
$f_{28}$  & & $2696.19$   & $0.17$   & $370.89$   & $0.024$   & $0.00939$   & $0.00052$   & $0.522$   & $0.018$   & $18.1$   & 16*$f_{\rm orb}$ \tabularnewline
$f_{23}$  & & $3033.26$   & $0.13$   & $329.68$   & $0.014$   & $0.01158$   & $0.00049$   & $0.670$   & $0.013$   & $23.7$   & 18*$f_{\rm orb}$ \tabularnewline
$f_{16}$  & & $3201.824$   & $0.081$   & $312.32$   & $0.0079$   & $0.01819$   & $0.00048$   & $0.9213$   & $0.0084$   & $38.1$   & 19*$f_{\rm orb}$ \tabularnewline
$f_{17}$  & & $3370.411$   & $0.092$   & $296.70$   & $0.0081$   & $0.01562$   & $0.00046$   & $0.2217$   & $0.0094$   & $33.7$   &  20*$f_{\rm orb}$ \tabularnewline
$f_{37}$  & & $3539.16$   & $0.22$   & $282.55$   & $0.017$   & $0.00660$   & $0.00046$   & $0.482$   & $0.022$   & $14.4$   & 21*$f_{\rm orb}$ \tabularnewline
$f_{54}$  & & $3706.50$   & $1.1$   & $269.80$   & $0.077$   & $0.00124$   & $0.00044$   & $0.54$   & $0.11$   & $2.8$   &  22*$f_{\rm orb}$ \tabularnewline
$f_{48}$  & & $4886.75$   & $0.53$   & $204.63$   & $0.022$   & $0.00244$   & $0.00042$   & $0.851$   & $0.055$   & $5.8$   & 29*$f_{\rm orb}$  \tabularnewline
$f_{45}$  & & $5392.15$   & $0.44$   & $185.45$   & $0.015$   & $0.00293$   & $0.00042$   & $0.188$   & $0.045$   & $7.1$   & 32*$f_{\rm orb}$ \tabularnewline
$f_{50}$  & & $5897.95$   & $0.58$   & $169.55$   & $0.017$   & $0.00216$   & $0.00040$   & $0.495$   & $0.060$   & $5.3$   & 35*$f_{\rm orb}$ \tabularnewline
$f_{47}$  & & $6740.91$   & $0.47$   & $148.35$   & $0.010$   & $0.00258$   & $0.00039$   & $0.869$   & $0.048$   & $6.6$   & 40*$f_{\rm orb}$ \tabularnewline
$f_{46}$  & & $7245.77$   & $0.42$   & $138.01$   & $0.0080$   & $0.00285$   & $0.00039$   & $0.279$   & $0.043$   & $7.3$   & 43*$f_{\rm orb}$ \tabularnewline
$f_{49}$  & & $7751.10$   & $0.52$   & $129.01$   & $0.0086$   & $0.00228$   & $0.00038$   & $0.618$   & $0.053$   & $6.0$   & 46*$f_{\rm orb}$ \tabularnewline
$f_{53}$  & & $9099.72$   & $0.64$   & $109.89$   & $0.0077$   & $0.00185$   & $0.00038$   & $0.304$   & $0.066$   & $4.8$   & 54*$f_{\rm orb}$ \tabularnewline
%$f_{51}$  & & $10350.28$   & $0.56$   & $96.61$   & $0.0052$   & $0.00213$   & $0.00038$   & $0.324$   & $0.058$   & $5.5$   &  \tabularnewline
%$f_{52}$  & & $10855.37$   & $0.59$   & $92.12$   & $0.0050$   & $0.00207$   & $0.00039$   & $0.695$   & $0.060$   & $5.3$   &  \tabularnewline

\end{longtable}

\end{document}